\pdfoutput=1
\documentclass[11pt]{article}

\usepackage[preprint]{acl}

\usepackage{times}
\usepackage{latexsym}
\usepackage{adjustbox}

\usepackage[T1]{fontenc}
\usepackage[utf8]{inputenc}

\usepackage{microtype}

\usepackage{inconsolata}

\usepackage{graphicx}

\usepackage{booktabs}
\usepackage{subcaption}
\usepackage{amsmath}
\usepackage{multirow}
\usepackage{multicol}
\usepackage{hyperref}
\usepackage{url}

\usepackage{siunitx} % for aligning results in talbes
\title{Robustness and Confounders in the Demographic Alignment of LLMs with Human Perceptions of Offensiveness}
\author{
    Shayan Alipour\textsuperscript{1}, Indira Sen\textsuperscript{2}, Mattia Samory\textsuperscript{1}, and {\bf Tanushree Mitra\textsuperscript{3}}\\
    \textsuperscript{1}~Sapienza University of Rome, 
    \textsuperscript{2}~University of Mannheim, 
    \textsuperscript{3}~University of Washington\\
    \texttt{shayan.alipour@uniroma1.it}, \texttt{indira.sen@uni-mannheim.de}, \\ \texttt{mattia.samory@uniroma1.it}, \texttt{tmitra@uw.edu }
}
\begin{document}
\maketitle
\begin{abstract}
Large language models (LLMs) are known to exhibit demographic biases, yet few studies systematically evaluate these biases across multiple datasets or account for confounding factors. In this work, we examine LLM alignment with human annotations in five offensive language datasets, comprising approximately 220K annotations. Our findings reveal that while demographic traits, particularly race, influence alignment, these effects are inconsistent across datasets and often entangled with other factors. Confounders---such as document difficulty, annotator sensitivity, and within-group agreement---account for more variation in alignment patterns than demographic traits alone. Specifically, alignment increases with higher annotator sensitivity and group agreement, while greater document difficulty corresponds to reduced alignment. Our results underscore the importance of multi-dataset analyses and confounder-aware methodologies in developing robust measures of demographic bias in LLMs. 
\end{abstract}

\section{Introduction}
 
% Commercial data annotation has been criticized for its lack of representativity of the broader population, over-reliance on few individual labelers, and the presence of low, 

A growing body of literature explores LLMs as a quick, cheap, and dependable alternative to human annotators ~\cite{chiang2023can,tornberg2023chatgpt,zhu2023can,gilardi2023chatgpt}. %The performance of LLMs has been found to match or even surpass that of human annotators for several annotation tasks. 
The eventuality of annotating data with LLMs is more than simple speculation: professional annotators already rely on LLMs to speed up their work~\cite{veselovsky2023prevalence}. Since good-quality data annotations are a primary concern for any machine-learning application, its essential to assess the quality of LLM-generated annotations. 

In particular, tasks that reflect annotators' subjective judgment and experiences, such as the perceived offensiveness of a message~\cite{davani2023disentangling}, raise the question of what kind of subjectivities are reflected in LLM-generated annotations. Recent research finds evidence of demographic bias, that is, systematic alignment between LLMs' annotations and those from select demographic groups of human annotators. If LLMs replicate the views of one demographic group over others, the resulting data applications risk perpetuating structural harms like marginalization of minority views. 

Although LLMs' demographic bias has been identified for a variety of subjective constructs including resume screening~\cite{wilson2024gender}, medicine~\cite{omiye2023large, zack2023coding}, political opinions~\cite{motoki2024more}, and offensiveness~\cite{sun2023aligning, santynlpositionality}, we know little about whether such bias holds beyond individual datasets. Moreover, many of these studies focus on different NLP tasks and datasets. For example, for offensiveness annotations, \citeauthor{sun2023aligning} find that LLMs align most with White and female annotators in the POPQUORN dataset, while~\citeauthor{schäfer2024demographicsllmsdefaultannotation} use the same dataset and do not find alignment between LLMs and women. Using a different dataset,~\citeauthor{santynlpositionality} find LLM alignment on offensiveness to be highest with Asian Americans. 
Thus, current findings are often contradictory due to the use of single datasets or idiosyncratic alignment measurement approaches, even if the same LLMs are studied. Understanding which demographic biases are consistent in LLM annotations is fundamental to tackling them. Our work fills this gap with a systematic study of LLM alignment on offensiveness labeling with different genders and ethnicities across five datasets, while also investigating factors beyond annotator demographics that might drive human-LLM misalignment.

First, we verify \textit{RQ0: to what extent LLMs can substitute human annotators in detecting offensive language}, to establish that LLMs' bias may not simply be attributed to low performance. Indeed, LLMs are strong performers: correlations with aggregate human labels are positive and significant---ranging from 0.4 and 0.8. Through permutation and bootstrapping tests, we show that LLMs surpass individual human annotators in three out of five datasets, while the performance in the remaining cases is comparable.

Next, we test
\textit{RQ1: which demographic biases are consistently reproduced in LLM-generated annotations across datasets.} Demographic biases exist within each dataset; however, most of these biases lack consistency. In our experiments we used two methods---rounded average and majority vote---to calculate the final label for a post. We found that only the bias between the White and Black demographics remained consistent when using the rounded average across all five datasets. Other differences between demographic pairs either appeared in only some datasets or even showed opposite results in different datasets.

% Demographic bias in favor of  White annotators (vs. Black or Asian) is robust across datasets and models. However, demographic bias toward other ethnicities and genders is inconsistent. 

We, therefore, explore \textit{RQ2: to what extent confounding factors explain demographic bias}, to understand how different characteristics of the datasets may lead to varying measures of demographic bias. %Building on Item-Response Theory scholarship, 
We consider three alternative hypotheses: HPa) how difficult documents to annotate may be assigned unevenly across demographics; HPb) how individual annotators may have idiosyncratic annotation preferences that dominate those of the demographic; and HPc), how annotators in a demographic sample may have varying levels of agreement, in turn affecting LLMs' alignment. We find them significantly related by modeling the relationship between demographic bias and confounders via logistic regression. Confounders explain a large fraction of the variance of LLM--human alignment, and thus can help us unpack cases when demographic bias is inconsistent.

\textbf{Overall contributions and novelty.} Unlike past work which has mainly focused on single datasets for assessing human-LLM alignment for data annotation, we investigate alignment between humans and LLMs for labeling offensive content across five different datasets. Depending on the dataset and contrary to past work~\cite{sun2023aligning,santynlpositionality},\textit{ we do not find consistent gender alignment effects or better alignment with Asian Americans}. While, we do find consistent patterns of better alignment with White people vs. Black people, demographic variables only drive a fraction of misalignment. We unpack this through a regression analysis with other confounding factors, like annotation difficulty. Finally, we contribute a harmonized, preprocessed, and merged version of five different NLP datasets modeling offensive language spanning 219,359 annotations, disaggregated by annotator groups that we make openly available for further community use.\footnote{\url{https://github.com/shayanalipour/llm-alignment-bias}}

% \url{https://github.com/TheLurkingLogger/llm-demographic-bias}

\section{Data}
To fully understand how LLM annotations align with human opinions, we leverage 5 datasets encoding annotators' perceptions of offensiveness---a construct that has been proven to vary according to annotators' sociodemographic characteristics. Specifically, we use the following datasets: 1) \textbf{``Annotator with Attitudes''} (AwA) dataset~\cite{sap2021annotators}, 2) UC Berkeley's \textbf{Measuring Hate Speech Corpus} (MHSC) ~\citep{kennedy2020constructing}, 3) \textbf{NLPositionality} (NLPos) dataset~\cite{santynlpositionality}, 4) \textbf{POPQUORN dataset} (POPQ)~\citep{pei2023annotator}, and 5) \textbf{Social Bias Inference Corpus} (SBIC)~\citep{sap2020social}.\footnote{Offensiveness is examined in three of the five datasets studied in this work (AwA, POPQ, and SBIC), while the remaining two datasets (NLPos and MHSC) focus on hate speech. Past research has looked into the association between offensiveness language and hate speech, concluding that both constructs, while not being the same, are often similarly perceived~\cite{davidson2017automated,founta2018large,fortuna2018survey}. Furthermore, offensive language can be considered a superset of hate speech, where the latter is offensive language targeting protected groups or minorities.} We use the annotator backgrounds in these datasets to consider the potential alignment between LLMs' responses and those of annotators in specific demographic groups (Table \ref{tab:dataset-demographics}). Following past research~\cite{sap2021annotators,sap2019risk}, we focus on annotators' ethnicity and gender as major factors in demographic alignment. Although age is a sociodemographic trait also present in all of the datasets included in this study, gender and ethnicity have the advantages that they are coded harmonically across datasets, and that their empirical distributions are such that they are likely to provide sufficient statistical power for analyses. More details about the datasets and their annotators can be found in the Appendix~\ref{sec:app_data}.
%Notably, gender distributions across datasets and within specific racial groups were evenly represented, allowing for more effective comparison. 

\begin{table}[htb]
\footnotesize
\centering
\setlength{\tabcolsep}{4pt}
\begin{tabular}[t]{lSSSSS}
\toprule
Demographic & AwA & MHSC & NLPos & POPQ & SBIC \\
\midrule
Man & 56.32 & 43.07 & 43.02 & 48.53 & 48.08 \\
Woman & 43.68 & 56.93 & 56.98 & 51.47 & 51.92 \\
\midrule
Asian & {-} & 5.95 & 20.92 & 7.85 & 6.73 \\
Black & 33.83 & 8.64 & 7.34 & 13.11 & 4.10 \\
Hispanic & {-} & 4.01 & 9.40 & {-} & 6.54 \\
White & 66.17 & 81.40 & 62.35 & 79.04 & 82.63 \\
\bottomrule
\end{tabular}
\caption{Demographic distributions across five datasets shown as percentages. %Note that some demographic data is missing for certain datasets (e.g., Hispanic demographic is not available for POPQ and AwA).
}
\label{tab:dataset-demographics}
\end{table}

\section{Methods}
In this section, we outline the models used for annotating data, explain our approach to prompting, and describe how we evaluate each of our research questions. 

\subsection{Models}
We conduct our experiments with two state-of-the-art models, GPT4o mini \citep{achiam2023gpt} and Gemini 1.5 Flash \citep{team2024gemini}, to identify consistent trends in detecting offensive and hateful language and explore any inter-model variations. Recent studies show that while GPT models perform well on hate speech detection \cite{huang2023chatgpt}, they may exhibit inherent biases, aligning more closely with certain demographics than others
%, which could influence the detection outcomes in sensitive task 
\citep{zack2024assessing,wang2023decodingtrust,tao2023auditing}. 
Similarly, Gemini 1.5 Flash which is a distilled version of Gemini 1.5 Pro, has been reported to achieve near parity with OpenAI's models across many benchmarks~\citep{team2024gemini}. 

%ChatGPT is the most studied LLM for its usability and suitability for text annotation tasks. While ChatGPT is the name for the end-user personal assistant tool, the specific model that powers this service and its configurations are not made public. For conducting our experiments, we use one of the latest OpenAI models, GPT-4o. 

% ChatGPT appears capable of detecting even hate speech that is implicit in nature, indicating that it is suitable for the experiments in this study \cite{huang2023chatgpt}. In particular, researchers have also shown that GPT-4 is capable of perpetuating biases and that it is more aligned with certain demographics than others \citep{zack2024assessing,wang2023decodingtrust,tao2023auditing}. 

% Gemini is a family of multimodal large language models developed by Google DeepMind. The model that we use, Gemini 1.5 Flash, is distilled from Gemini 1.5 Pro, with a context length above 2 million. Gemini1.5 has been shown to be highly performant for several different tasks, on par with GPT4~\cite{..}.

% Despite the high-performance of OpenAI models across a variety of tasks, their proprietary nature potentially affects their viability as synthetic annotators. Although the details are not disclosed by OpenAI, it is evident that safety guardrails have been put in place to prevent the GPT models from generating questionable content, which results in the models abstaining from providing output. The lack of details on the safety guardrails is further exacerbated by their periodic updates, which significantly modify their behavior \citep{chen2023chatgpt}.

\subsection{Prompting Strategies}\label{sec:prompting}

We designed a scoring system that matches the questions asked to human annotators across various datasets for evaluating offensive or toxic content. %In the POPQ and AwA datasets, the models used a 5-point scale, where 1 meant "not toxic" and 5 meant "extremely toxic." For the MHSC and NLPos datasets, we used a 3-point scale with the options: 1 (Yes), 2 (Unclear), and 3 (No) for MHSC, and 1 (Yes), 2 (Not sure), and 3 (No) for NLPos. In the SBIC dataset, the valid options were: 1 (Yes, this could be offensive), 2 (Maybe, I'm not sure), 3 (No, this is harmless), and 4 (I don't understand the post). 
Following the approach in \cite{wei2022chain}, we required the model to not only rate the comments but also justify its scores. The wording of the prompts was the same as that used for human annotators, with an additional instruction: ``Begin your response by mentioning one of the valid options, then provide a concise explanation for your rating." Our prompts are included in Table ~\ref{tab:prompts} in the Appendix. We used a regular expression to extract the final labels from the models' responses. %The distribution of labels assigned by both annotators and models across different datasets and rating levels is available in SI Table 7. 

\subsection{RQ0:  LLMs as Annotators}
To evaluate how effectively LLMs detect uncivil language, we compared model-generated labels with human annotators’ labels across multiple datasets. We calculated ground truth values using two methods: a rounded average of the labels and the majority label. The main body focuses on the rounded average method, while the majority-label results are detailed in the supplementary information. We calculated the Pearson correlation coefficients between the model and human labels, performed t-tests to assess statistical significance, and calculated 95\% confidence intervals (CI) using bootstrapping with 1,000 samples. To further test the robustness of the correlations, we ran a permutation test by shuffling demographic information and recalculating correlations for 1,000 iterations.

\begin{figure}
  \includegraphics[width=\linewidth]{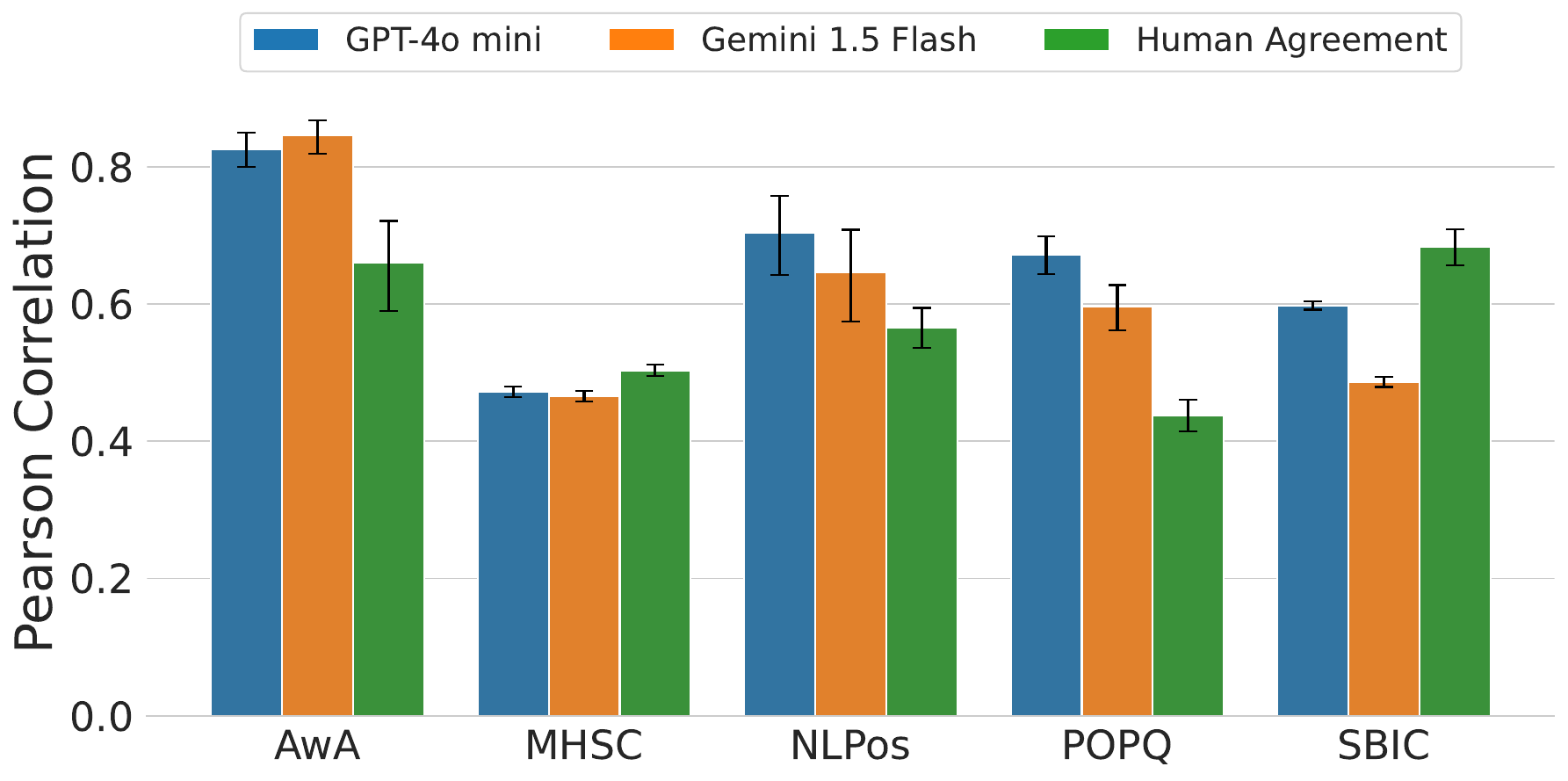}
  \caption {Comparison of model correlations with human annotators against human agreement (individual annotators with their peers) which highlights how well models align with human judgment. 
  % The ground truth for each post is calculated using the average of all annotators' labels. 
  % For human agreement, correlations are measured by leaving out one annotator and comparing their labels to the ground truth from the remaining annotators. Error bars represent 95\% confidence intervals.
  }

  % , computed by bootstrapping: model correlations are bootstrapped directly, while human agreement is bootstrapped by sampling the mean values from individual annotators’ correlations
  
    \label{fig:average-human-vs-model}
\end{figure}

To measure agreement among human annotators, we used a leave-one-out approach. For each annotator, we excluded their label from each post and recalculated the ground truth from the remaining labels. This was repeated for every post that the annotator labeled. We then measured the correlation between the annotator’s labels and the recalculated ground truth. By averaging these correlations across all annotators, we determined overall human agreement.
To evaluate the robustness, we applied bootstrapping (1,000 samples) to this correlation distribution to estimate 95\% confidence intervals.

\subsection{RQ1: Demographic Bias Robustness}
To analyze demographic biases in LLM-generated annotations, we calculate the ground truth for each demographic by filtering the annotations for that group and aggregating them per post. We then compute the Pearson correlation $r$ between the model’s predictions and the demographic-specific ground truth. We also apply the aforementioned robustness checks, including t-tests, confidence interval estimation using bootstrapping, and permutation tests on the demographic labels. 

To assess whether the model consistently aligns better with one demographic than another, we use Steiger's $Z$ test~\cite{steiger1980tests, hoerger2013zh} to determine if the difference between correlations $\Delta r$ is statistically significant. Steiger's $Z$ test compares correlations that share a common variable, in this case, the model’s predictions. To account for multiple comparisons, we adjust $p$-values using the Holm-Bonferroni correction. Additionally, we compute confidence intervals for the difference in correlations using bootstrapping. In particular, we measured $\Delta r = r(P, D_1) - r(P, D_2)$ where $P$ represents the model's predictions, $D_1$ and $D_2$ are two demographic groups. This involves resampling the annotations for each demographic pair and recalculating the difference over 1,000 iterations. If the 95\% CI for the bootstrapped distribution includes zero, this suggests that the observed difference in correlations may be due to random variation in the sample distribution.

\begin{figure*}[htb!]
  \centering
  \begin{subfigure}[t]{0.48\textwidth}
    \centering
    \caption{Gender}
    \includegraphics[width=\linewidth]{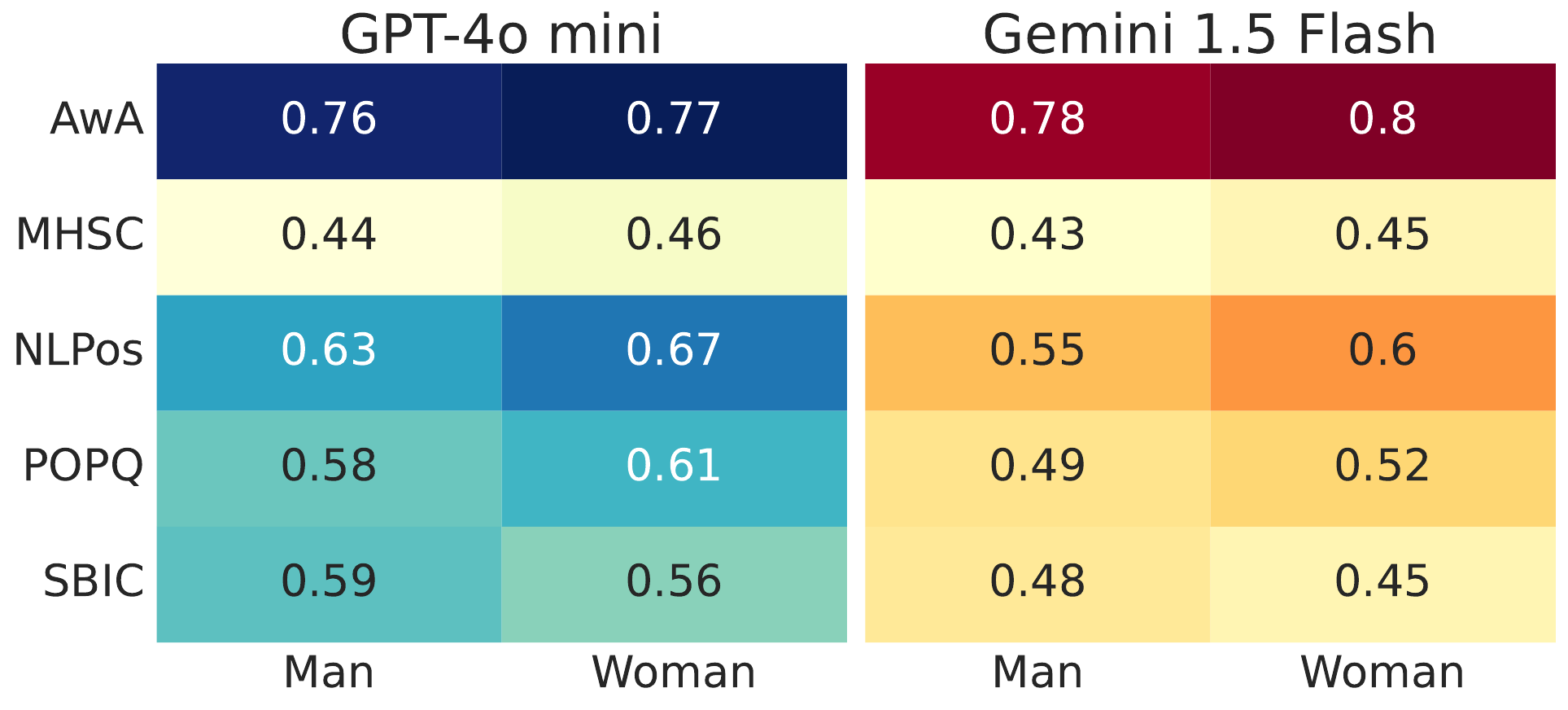}
  \end{subfigure}%
  \hfill
  \begin{subfigure}[t]{0.48\textwidth}
    \centering
    \caption{Ethnicity}
    \includegraphics[width=\linewidth]{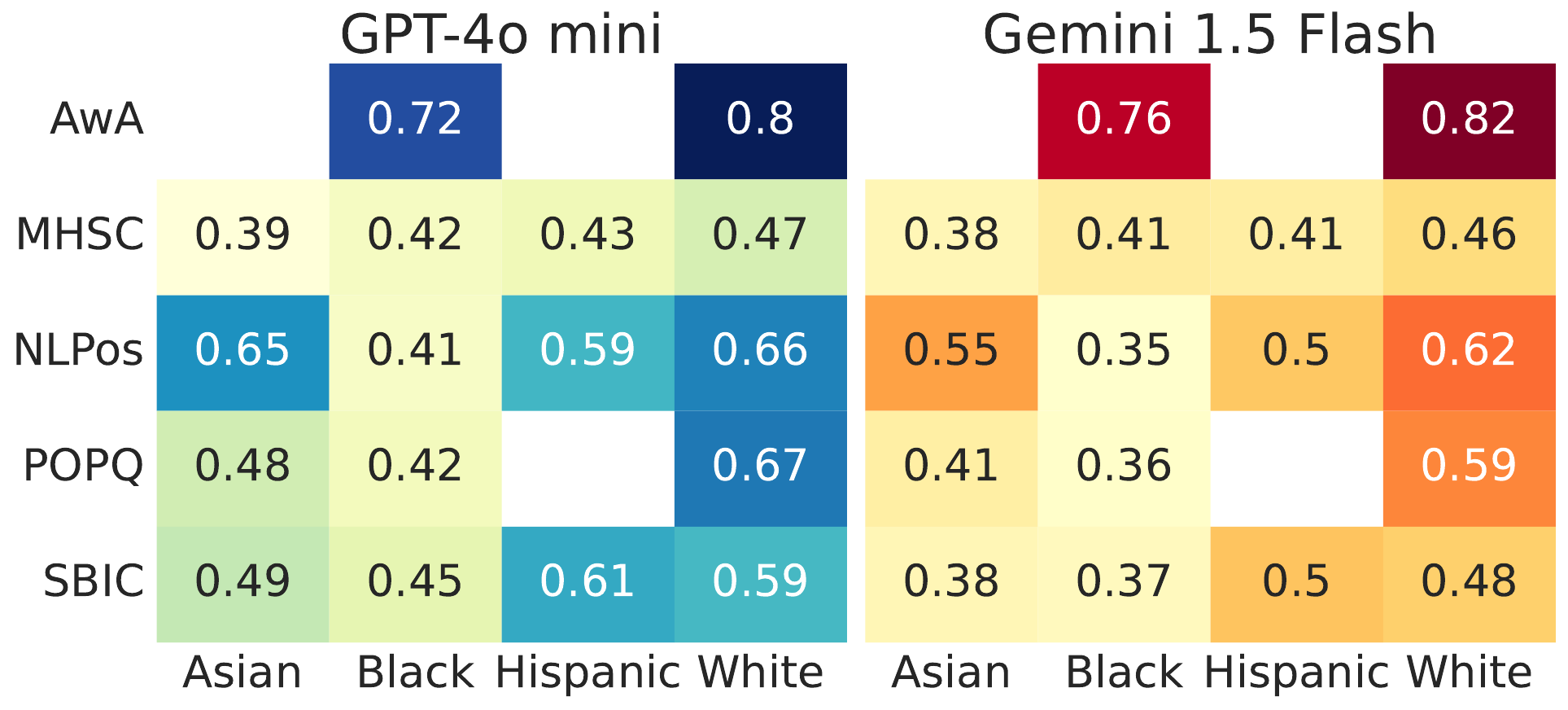}
  \end{subfigure}
  
  \caption{
    Pearson correlation coefficients between model outputs and human annotator labels, broken down by gender (a) and ethnicity (b) across five datasets. The ground truth for each post is determined by averaging the labels from annotators belonging to the target demographic. Darker shades indicate stronger correlations. Confidence intervals and p-values for statistical significance are reported in Table \ref{tab:average-demo-model-corr} in the Appendix.
  }
  \label{fig:gender_and_race_corrs}
\end{figure*}

\subsection{RQ2: Demographic Bias Confounders}
We consider alternative hypotheses for the alignment between the LLMs and humans, beyond demographic bias. We use individual annotations as observations and operationalize \textbf{alignment} as an indicator variable set to 1 when LLM and human annotations coincide. While considering exact alignment does not capture the direction and magnitude of the differences in annotations, this operalization is practically useful since alignment is high overall (\textasciitilde 60\% of annotations align perfectly). We model alignment 
via logistic regression with annotators' demographic traits as independent variables. Next, we develop additional hypotheses for factors that may confound demographic alignment, which we include as additional independent variables.
        \paragraph{HPa: difficulty.} Documents that are difficult to annotate may be assigned unevenly across demographics, which may in turn negatively affect LLM's alignment. We measure difficulty as the negative Kullback-Leibler divergence between a document's labels and the uniform distribution. Intuitively, the more diverse the annotators' labels, the more difficult the document is to annotate.  
        \paragraph{HPb: sensitivity.} Since few annotators partake in most annotation tasks, some annotators contribute more labels than others, and the representation of demographic traits is unequal, individual annotator factors may dominate the apparent demographic biases. At its simplest, some annotators may systematically label documents as more offensive than other annotators, irrespective of their demographic or a document's aggregate label. To measure sensitivity, we rank annotators of a document based on their labels. The higher the annotators' rank, the more likely they align with LLMs tuned to discourage offensive content.
        \paragraph{HPc: agreement.} Although alignment is typically measured with a whole demographic group of annotators, different groups may have varying levels of internal agreement. When a group internally disagrees, alignment with it is more complex (and arguably, less meaningful). We operationalize agreement as the negative absolute difference between the individual annotators' labels and the average label of their demographic groups, therefore computing distinct agreement values for gender and ethnicity. The more annotators behave similarly to their reference group, the higher the agreement.

In addition to confounders, we include the document's offensiveness \textbf{label} as a control variable to account for skews in the LLMs' annotations, since we aim at measuring whether LLMs replicate demographics' annotations of individual documents rather than generic similarity in label distributions in a dataset. LLMs may prefer to label documents as offensive, given their terms of service and training for application in general, safe-for-work contexts. We also control for dataset-specific levels of the dependent and independent variables by including corresponding intercepts. We center all binary variables and standardize confounders and documents' labels within each dataset by centering them and dividing their values by two standard deviations to make their scales comparable and interpretable \cite{gelman2008scaling}.

\section{Results}
In this section, we address three questions about the capabilities of LLMs in simulating human judgment when annotating offensive language. First (\textbf{RQ0}), we examine the extent to which LLMs can accurately replicate human annotations overall. Second (\textbf{RQ1}), we investigate the alignment between these models and the annotations of sociodemographic subgroups of annotators and whether these alignments are consistent across datasets. Third (\textbf{RQ2}), we model the alignment behavior by considering potential confounding factors.

\subsection{RQ0: Viability of LLMs as Annotators}\label{sec:rq0}

Our results demonstrate that LLMs closely mirror human annotations. Figure~\ref{fig:average-human-vs-model} shows that the correlations between LLM labels and ground truth labels are strong, positive, and significant, measured using both rounded averages and majority votes.\footnote{Figure~\ref{fig:majority-human-vs-model} in the Appendix shows the results for the majority votes, which display a similar pattern.} Specifically, the overall correlations range from $r = 0.47$ to $r = 0.85$ across the datasets and models, indicating the models' strong alignment with human consensus. To put model performance into perspective, we compare them with the performance that individual human annotators achieve compared to the remaining annotators. In three datasets---AwA, POPQ, and NLPos---LLMs surpass individual human annotators. In the other two datasets (SBIC and MHSC), LLMs perform competitively.

\subsection{RQ1: Robustness of Demographic Bias}\label{sec:rq1}
Figure~\ref{fig:gender_and_race_corrs} shows the correlation values $r$ between LLM labels and the labels of annotators from each demographic group. In Figure~\ref{fig:gender_and_race_corrs}.a both GPT-4o and Gemini show similar patterns, aligning more closely with annotations from women in the AwA, MHSC, NLPos, and POPQ datasets, while aligning better with men's annotations in SBIC. For ethnicity, Figure~\ref{fig:gender_and_race_corrs}.b shows that both models align better with the White demographic in four datasets, except in SBIC, where they achieve higher $r$ with the Hispanic demographic. The Hispanic demographic appears in two other datasets---MHSC and NLPos---where they rank third and second out of four ethnic groups, respectively. Across most datasets, the Black demographic generally shows the lowest correlations, except in MHSC, where it surpasses the Asian demographic. These correlation values, as summarized in Table~\ref{tab:average-demo-model-corr}, are significant and not due to chance, as confirmed by our permutation tests (see Figures~\ref{fig:permutation-average} and~\ref{fig:permutation-majority} in the Appendix). 

\begin{figure*}[t] 
  \centering
  \begin{subfigure}[t]{0.48\textwidth}
    \centering
    \caption{GPT-4o mini}
    \includegraphics[width=\linewidth]{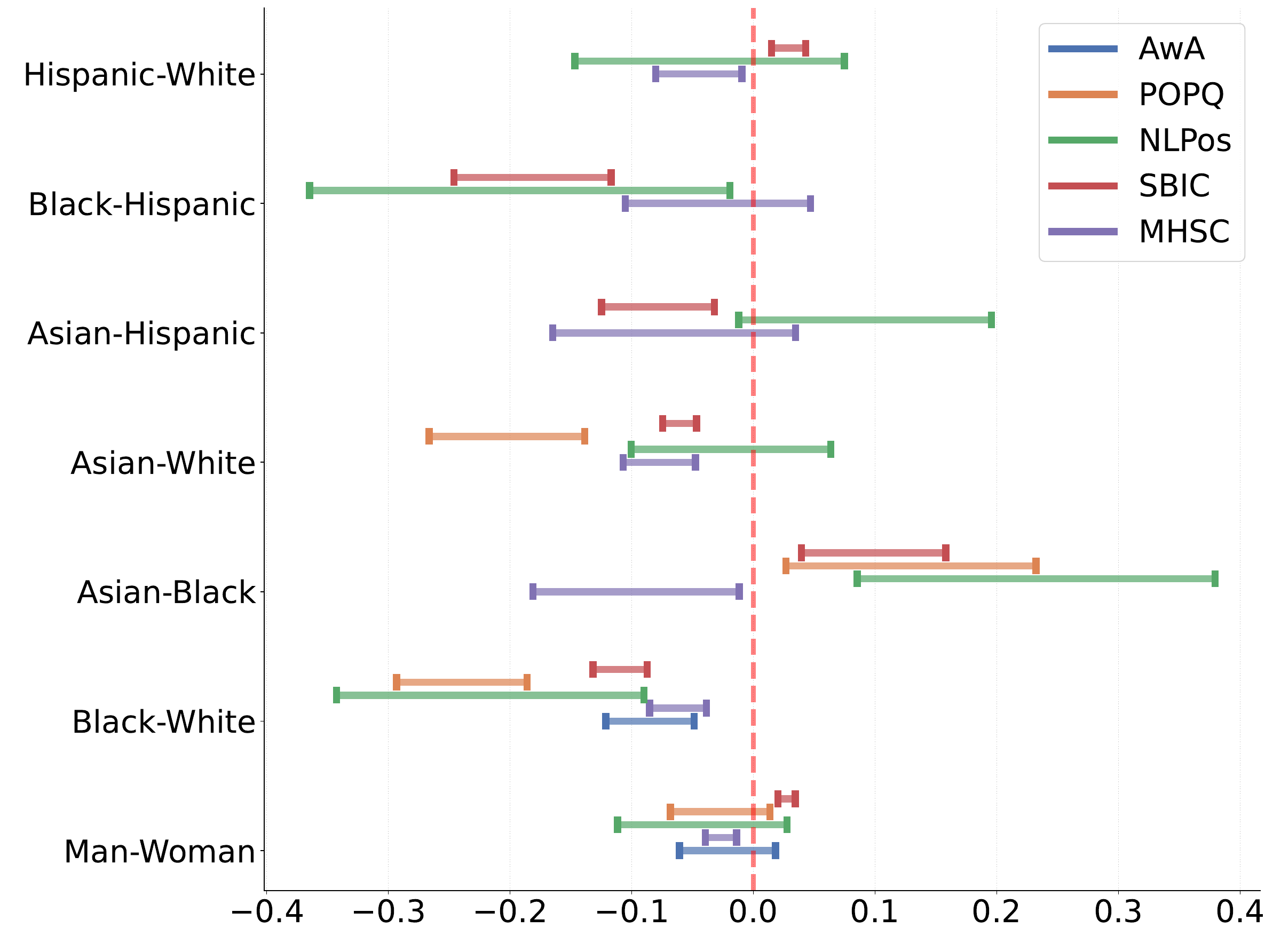}
  \end{subfigure}%
  \hfill 
  \begin{subfigure}[t]{0.48\textwidth}
    \centering
    \caption{Gemini 1.5 Flash}
    \includegraphics[width=\linewidth]{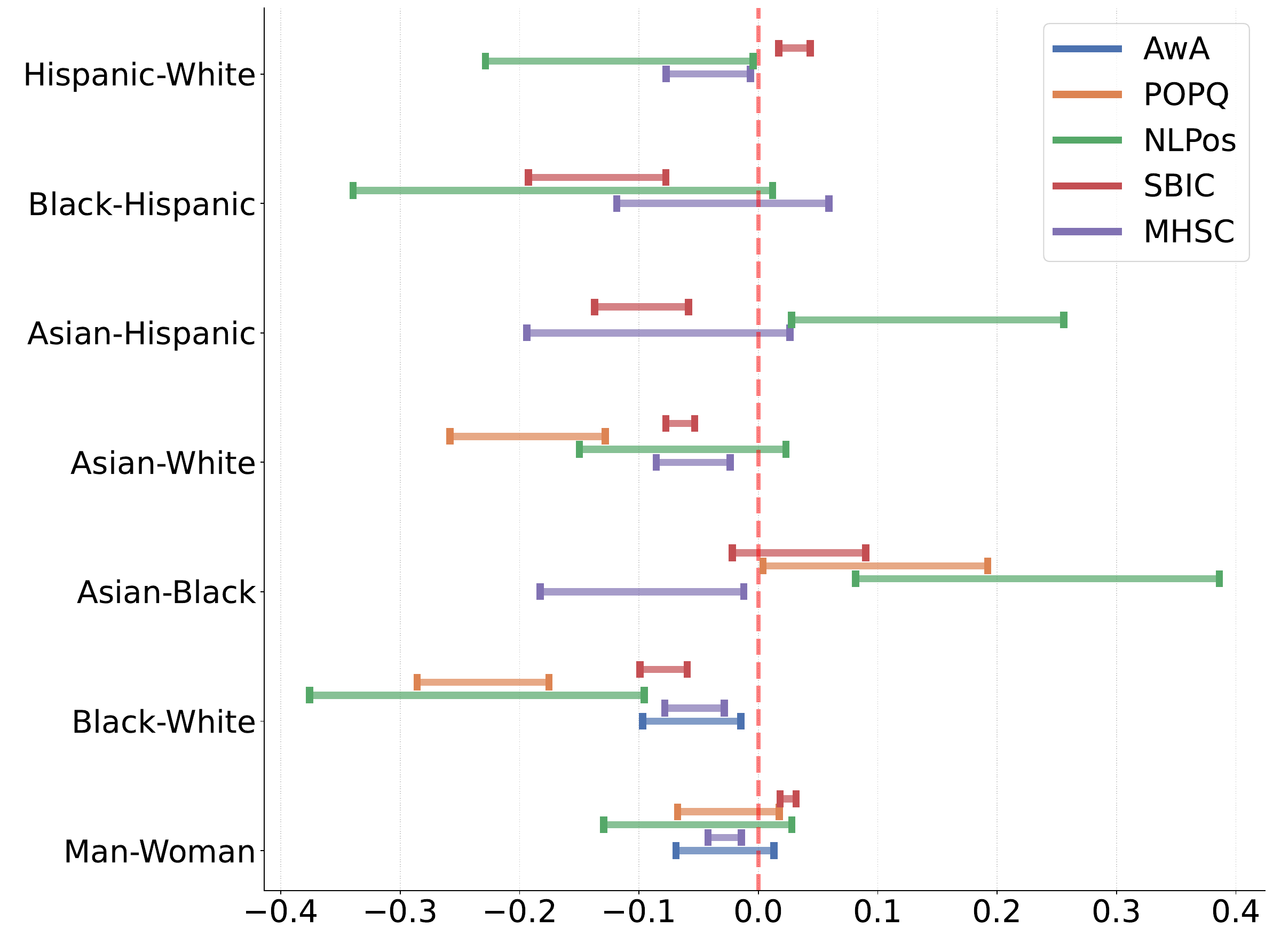}
  \end{subfigure}
  
  \caption{The 95\% confidence intervals (CI) for the difference in correlation between the model's predictions and two demographic groups, computed as: $\Delta r = r(P, D_1) - r(P, D_2)$, where $P$ represents the model's predictions, and $D_1$ and $D_2$ are two demographic groups. Ground truth for each post is determined by averaging the labels from annotators in the target demographic. The intervals are derived from 1,000 bootstrap samples. If the CI includes zero, the difference is not statistically significant. See Table~\ref{tab:average-corr-diff} in the Appendix for further details.}
  \label{fig:average_corr_diff}
\end{figure*}

However, echoing~\citeauthor{movva2024annotation}, statistical significance alone doesn’t guarantee consistent alignment across demographic groups. To assess whether these differences are robust and meaningful, we examine the correlation differences using  Steiger’s $Z$ test and bootstrapping. Without this additional analysis, there's a risk of over-interpreting the correlations, which might reflect dataset-specific variations rather than true alignment. Figure~\ref{fig:average_corr_diff} displays the results for demographic pair comparisons when using the rounded average aggregation method. Some comparisons reveal no significant or robust differences. For example, in the POPQ dataset, the correlation difference between men and women changes sign depending on the sample. We observe that the models align slightly better with the Hispanic demographic than with the White demographic in SBIC, but this trend reverses in MHSC and NLPos. The only consistent finding across all datasets is the Black-White pair, where models show lower correlations with the Black people.

Results from Figure~\ref{fig:average_corr_diff} highlight a general lack of robustness indicating that demographic annotations are influenced by many factors beyond demographic identity, such as individual interpretation or dataset composition.\footnote{Results for the majority vote are similar and reported in the Appendix (Figure~\ref{fig:majority_corr_diff})} While the correlation values $r$ between models and annotator groups are statistically significant, testing the robustness of correlation differences shows that demographic identity alone does not consistently explain the observed variance in model alignment.
%In other words, a model’s alignment with one group over another may not be stable across all situations.

\subsection{RQ2: Significant Confounders of Demographic Alignment}\label{sec:rq2}

We now explore potential confounders that may account for inconsistencies in RQ1. Table \ref{tab:confounders} shows the summary statistics of the logistic regressions of the alignment between LLM and human labels (1 only if they use the same label). The first regression (on the left in the table) explains the relationship between LLM alignment and the annotator's gender and ethnicity, controlling for per-dataset differences. As for our previous analyses, alignment is higher for White than for Black annotators, with marginally significant differences between women and men. This demographic-only model explains little variance in the LLM-human alignment (pseudo-$R^2$ = 0.015). The second model (on the right in the table) attempts to explain the remaining variance by accounting for confounding factors. Indeed, modeling confounders substantially improves model fit (pseudo-$R^2$ = 0.213). All confounders show statistically significant coefficients and follow their hypothesized behavior (controlling for annotators' demographics). 

First, the more difficult it is to label a document, i.e., the more diversity in the annotators' labels regardless of their demographics, the lower the alignment with the LLM. Conversely, the higher the agreement between annotators of the same demographic on a document, the higher the likelihood of alignment. Additionally, the more sensitive one annotator is to offensiveness compared to other annotators of the same document, the higher the alignment, which remains true even when controlling for the overall label of the document, irrespective of the annotators' demographics. In fact, the largest coefficients are associated with confounders at the level of the document---its label and annotation difficulty---which do not directly model the preference of individuals or groups of annotators. 

Yet, these confounders do not fully mediate demographic alignment: when explicitly modeling them, we see increased rather than decreased significance of demographic coefficients, e.g., LLMs significantly align with male annotators in the confounder-aware model but not in the demographics-only model. Additionally, even with the confounders, the overall explanation for the variance in annotations is moderate (pseudo-$R^2$ = 0.213), indicating that other hidden confounders need to be accounted for to explain (mis)alignment, e.g., social media usage or attitudes towards free speech~\cite{fleisig2023majority,sap2021annotators}.

In summary, we find that \textbf{LLMs' demographic bias is at least partially explained by confounding factors}. Especially, LLMs' overall tendency to rate documents as offensive matches demographics that are assigned more, more clearly offensive documents and/or include more sensitive, mutually agreeing individual annotators. Since the procedures of annotator recruitment and document assignment for a demographic vary between annotation tasks, LLMs may appear biased toward different demographics in the resulting datasets. Thus, by resorting to factors at the document, individual annotator, and annotator sample levels, \textbf{we can reconcile inconsistencies in the demographic bias observed across datasets}---that would otherwise contradict the assumption that LLMs replicate certain demographics' annotations. 

\begin{table}[!htbp] \centering\sffamily\small
\begin{tabular}{@{\extracolsep{5pt}}lSS} 
% \\[-1.8ex]\hline
% \hline \\[-1.8ex]
% & \multicolumn{2}{c}{\textit{dep. var.: alignment}} \
% \cr \cline{1-3}
\\[-1.8ex] {D.V.: alignment} & {model 1} & {model 2} \\
\hline \\[-1.8ex]
 dataset=popq & -0.320$^{***}$ & -0.410$^{***}$ \\
% & (0.040) & (0.046) \\
 dataset=nlpos & 1.075$^{***}$ & 1.547$^{***}$ \\
% & (0.048) & (0.058) \\
 dataset=sbic & 0.802$^{***}$ & 1.018$^{***}$ \\
% & (0.036) & (0.042) \\
 dataset=mhsc & 0.843$^{***}$ & 1.055$^{***}$ \\
% & (0.036) & (0.042) \\
 gender=woman & -0.021$^{*}$ & -0.036$^{***}$ \\
% & (0.009) & (0.010) \\
 ethnicity=asian & -0.175$^{***}$ & -0.127$^{***}$ \\
% & (0.018) & (0.021) \\
 ethnicity=black & -0.071$^{***}$ & -0.118$^{***}$ \\
% & (0.018) & (0.021) \\
 ethnicity=hispanic & 0.049$^{*}$ & -0.045$^{}$ \\
% & (0.020) & (0.024) \\
 difficulty & & -1.747$^{***}$ \\
% & & (0.015) \\
 sensitivity & & 0.499$^{***}$ \\
% & & (0.011) \\
 agreement\textsubscript{ethnicity} & & 0.251$^{***}$ \\
% & & (0.015) \\
 agreement\textsubscript{gender} & & 0.208$^{***}$ \\
% & & (0.014) \\
 label & & 1.788$^{***}$ \\
% & & (0.013) \\
 intercept & 0.439$^{***}$ & 0.545$^{***}$ \\
% & (0.004) & (0.005) \\
\hline \\[-1.8ex]
 observations & {219359} & {219359} \\
 pseudo $R^2$ & 0.015 & 0.213 \\
 % \hline
\hline \\[-1.8ex]
 & \multicolumn{2}{r}{$^{*}$p$<$0.05; $^{**}$p$<$0.01; $^{***}$p$<$0.001} \\
\end{tabular}
\caption{Logistic regression of LLM--human alignment. Model 1 (left) explains whether an LLM chooses the same label as a human annotator by regressing over the annotator's gender (vs. man as the reference level ), ethnicity (vs. White), and the annotated document's dataset (vs. AwA), encoded as indicator variables. Model 2 (right) additionally accounts for potential confounders: the document's difficulty, the annotator's sensitivity, and the agreement of the annotator with other annotators of the same gender and ethnicity, as well as the annotator's label as a control variable to account for the LLM's overall label skew. %The additional regressors in model 2 are standardized. We included only documents with at least one annotator per gender and ethnicity.
}\label{tab:confounders}
\end{table}

\section{Related Work}

\begin{table*}[h!]
\centering
\begin{adjustbox}{max width=\textwidth}
\begin{tabular}{p{2cm}p{4cm}p{4cm}p{3cm}p{4cm}p{3cm}p{6cm}}
\toprule
\textbf{Paper} & \textbf{Models Used} & \textbf{Datasets} & \textbf{Task / Construct} & \textbf{Demographic Variables} & \textbf{Demographic / Default Prompting} & \textbf{Findings for Default Prompting} \\
\midrule
\citet{giorgi2024human} & Llama-3, Phi-3, SOLAR-10, Starling-LM & 1 (Measuring Hate Speech) & hate speech & gender, race, age, education, religion, sexuality, ideology & demographic & N/A \\
\citet{hu-collier-2024-quantifying} & GPT4, GPT3.5, LlaMa-2, Tulu & several & sentiment, stance, offensive language, hate speech, social acceptability & gender, race, location, education, native language, age, religion, political leaning & demographic & N/A \\
\citet{sun2023aligning} & FLAN-T5-XXL, FLAN-UL2, GPT-3.5, GPT-4 & 1 (POPQUORN) & offensiveness, politeness & gender, race & both & better alignment with White people and women \\
\citet{schäfer2024demographicsllmsdefaultannotation} & GPT4o, Claude & 1 (POPQUORN) & offensiveness, politeness & gender, race, age, education, occupation & both & better alignment with White people compared to Black people \\
\citet{beck2023not} & GPT-3, T5, OPT, Pythia & several & sentiment, stance, offensive language, hate speech & gender, race, age, education, political leaning & both & do not test or report alignment with the default prompt, i.e., the one without sociodemographic info. \\
\citet{santynlpositionality} & GPT4 & 1 (NLPositionality) & hate speech, social acceptability & gender, race, location, education, native language, age, religion & default & for hate speech: better alignment with Asian-Americans than White \\
\citet{movva2024annotation} & GPT4 & 1 (DICES) & safety & gender, race & default & no clear demographic alignment \\
\midrule
Ours & GPT4o, Gemini, LLaMa & several & offensiveness, hate speech & gender, race & default & better alignment with White people compared to Black people \\
\bottomrule
\end{tabular}
\end{adjustbox}
\caption{Summary of recent research on LLMs' annotations and their demographic alignment with humans.}
\label{tab:related_work}
\end{table*}

Our work assess one aspect of human-LLM alignment within the context of automated labeling --- demographic alignment. Therefore, it lies at the intersection of using LLMs for annotation, annotation subjectivity, and demographic bias in LLMs.

\subsection{Using LLMs for Data Annotation}

While pre-trained language models like BERT and RoBERTa have been used widely for content analysis~\citep{gonzalez2020comparing}, recent work has focused on using generative LLMs, such as Flan-T5 and GPT for data labeling for various social constructs like offensive language~\citep{zampieri2023offenseval}, stance detection~\citep{aiyappa2024benchmarking}, hate speech~\citep{huang2023chatgpt}, and framing~\citep{gilardi2023chatgpt}. However, there is contention about the quality of LLM-generated annotations. %human annotations are typically considered the gold standard. 
Since ChatGPT's release, some studies have claimed it outperforms human annotators~\citep{gilardi2023chatgpt,wu2023large,chiang2023can,tornberg2023chatgpt,zhu2023can}, while others find LLMs do not reach human-level performance~\citep{kristensen2023chatbots}.
% and noted its widespread use in crowd work~\citep{veselovsky2023prevalence}. 
% However, other research argues that Large Language Models like ChatGPT haven't yet reached human-level performance and compare similarly to earlier non-generative models like BERT~\citep{kristensen2023chatbots,ziems2024can}.

% Nonetheless, several researchers have looked into the performance of LLMs, mainly commercial models like GPT3.5 and GPT4 but also open-source ones like Mistral, Flan-T5, and LLama, for labeling content for a variety of tasks, e.g., stance detection~\citep{aiyappa2024benchmarking}, hate speech~\citep{huang2023chatgpt}, and framing~\citep{gilardi2023chatgpt}. Our findings add clarity to the diverse results seen in this area by highlighting specific factors that could impact how well LLMs align with human labelers.

% how is this paper positioned wrt previous work in this section
\subsection{Subjectivity and Demographic Factors in Human Annotations}

Data annotations from multiple annotators, collected for training supervised classification models, have traditionally been aggregated to a single ground-truth label for each instance, usually based on majority aggregation. 
%following the ``wisdom of the crowd'' paradigm---i.e., the assumption that each annotation is a noisy measure of a single, true label and that therefore an aggregation of multiple annotations is a better estimate of such true label. 
Yet, annotation is often an interpretive task that depends on the annotator's positionality, social situation, and lived experiences~\citep{paullada2021data,santynlpositionality}, which challenges the assumption of the existence of such a single true label. Recent work in NLP and ML has recognized that the typical majority aggregation can squash the opinions and views of minoritized and marginalized populations
%. Consequently, automated classifiers trained on data aggregated by majority labeling would only align with the dominant demographics in the annotator pool
~\citep{davani2022dealing,davani2023disentangling}. This is particularly pertinent for subjective tasks such as detecting offensive, abusive, or toxic content. Here, an annotator's demographic identity~\citep{al2020identifying}, attitudes~\citep{sap2021annotators}, personal experiences~\citep{sang2022origin}, and their combination impact annotator perception of toxicity. Furthermore, researchers also recommend explicitly factoring in dimensions that would lead to disagreement \textit{before} the annotation task to reduce haphazard annotation distributions~\cite{fleisig2024perspectivist}.% This has been described as ``Strong Perspectivism''.\footnote{see the Perspectivist Data Manifesto at \url{https://pdai.info/}}

% Given the role sociodemographic identity plays in human annotators' perceptions of offensive content, we assess if LLM annotations mimic certain identities more than others. Using a perspectivist outlook, We build on this body of work to evaluate the performance of LLMs as annotators, and investigate whether LLM-generated annotations reproduce the annotation patterns of people of various demographic backgrounds. 

% how is this paper positioned wrt previous work in this section
% \subsection{Issues with Automated Annotations}
% how is this paper positioned wrt previous work in this section
\subsection{Demographic Alignment of LLMs for Annotation}
% While LLMs excel at tasks requiring advanced linguistic competence, r
%Researchers have found that LLMs appear to reproduce the views of certain demographics, particularly people residing in the Global North~\citep{santurkar2023whose}. 
%While this raises questions about the representativity of LLMs, these studies usually confront LLMs with survey questions or personality inventories. What remains unclear is how these demographic misrepresentations affect LLM behavior for downstream usage, including their use as data annotators and the repercussions of using LLM-annotated data for training machine learning models~\citep{alemohammad2023self}.

% Building on findings of sociodemographic biases in LLMs when it comes to opinions~\cite{santurkar2023whose}, 
A growing body of work looks more specifically at the sociodemographic alignment of LLMs and human  annotators for the task of content analysis~\cite{sun2023aligning,santynlpositionality}, \textit{inter alia}. Table~\ref{tab:related_work} summarizes and compares this work to most related past and concurrent work on this, focusing on the different types of LLMs studies, the different NLP tasks, as well as on how many datasets. More importantly, we differentiate between \textit{default} vs. \textit{sociodemographic} prompting, where the latter includes demographic variables in the input prompt to LLMs in order to improve alignment. However, in this work, we focus on the default prompts without sociodemographic signals to first systematically investigate the default alignment of LLMs across several datasets, and thereby establish convergent validity~\cite{cunningham2001implicit}. 

Papers assessing default prompting are the closest to our current research; they mainly do so on single, but different datsets, leading to contradictory findings about demographic alignment, even for the same or related NLP tasks like offensiveness and hate speech. \citeauthor{sun2023aligning} look into the alignment between human annotators and LLM annotations for offensiveness and politeness, finding that LLMs align best with White women. \citeauthor{schäfer2024demographicsllmsdefaultannotation}use the same dataset, but find that LLMs align better with White people, but not women. Differing from both of these, ~\citeauthor{santynlpositionality} find that for hate speech detection, models align best with Asian Americans. Our work unpacks these seemingly contradictory findings by conducting a systematic study across several datasets, accounting for variance, and going beyond demographic variables. % Finally, \cite{giorgi2024human} assess if humans and LLMs show similar biases based on the target of hate speech in the instance to be labeled. 

\section{Discussion and Conclusions}\label{sec:discussion}

\textbf{Summary of findings. }Our findings corroborate some of the previous results on demographic biases in LLMs' offensiveness ratings. Specifically and in line with~\citeauthor{sun2023aligning} and~\citeauthor{schäfer2024demographicsllmsdefaultannotation}, LLMs consistently align better with White annotators vs. Black. This bias replicates across five datasets and is measurable even when accounting for several confounders. However, our systematic analysis offer a more nuanced picture compared to single-dataset studies from past work. Apparent demographic biases such as those based on annotators' gender~\cite{sun2023aligning} or the Asian American demographic~\cite{santynlpositionality}, are statistically significant but contradictory in different datasets. We show that LLM--human alignment may appear, unduly, as being associated with the annotators' demographic, when in fact it is confounded, if not explained by several factors that accrue in groups of annotators who share a demographic trait in a particular dataset but not in others.

\textbf{Implications for Demographic Alignment in Data Annotation. } The consistently worse alignment with Black people highlights inherent biases in the use of LLMs for detecting offensiveness. Therefore, we echo the warnings raised in related literature about the potential negative consequences of using LLM-generated annotations without first understanding whose points of view they may reinforce or neglect. However, given the lack of consistent alignment with any other demographic factor, we caution against narratives that anthropomorphize LLMs and essentialize annotators. % Instead, the complexity of our results warrants a critical approach when analyzing model performance and explaining unexpected, undesirable, or potentially harmful outputs. The expected reward for such admittedly onerous analysis is more effective strategies for model improvement, including enhanced guardrailing and debiasing techniques.

\textbf{Recommendations for Measuring Misalignment. }Our results show that measuring demographic biases in a single dataset may produce unreliable results. Systematic benchmarking across multiple datasets, coupled with replication and meta-analytical studies, is essential to support general claims about demographic bias. Moreover, including confounding variables in analyses is crucial to distinguish between demographic bias and other sources of LLM--human alignment; established frameworks like Item-Response Theory, which guide the hypotheses about confounders in the present work, offer valuable best practices.

The heterogeneity of results also underscores the need for datasets with greater demographic representation, more redundant labeling, and increased data diversity~\cite{fleisig2024perspectivist}. While acknowledging the practical and financial constraints of building such datasets, our research provides insights into strategic approaches for enhancing data quality. Specifically, we emphasize the importance of accounting for confounders, such as document difficulty and the sensitivity of individual annotators, in their interplay with annotator demographics. Since, at present, such confounders only emerge at the end of the annotation process, we see an opportunity for developing annotation solutions that dynamically adjust annotator recruitment and document assignment in response to emerging patterns. 

\section{Limitations}

% Our study's limitations include the varied prevalence of demographics within our data. While more prevalent demographics, like White women, might not directly influence model alignment, their internal consistency could simplify alignment. Additionally, specific challenges arise from the potential bias in document assignments to different demographics. For instance, if very few Asian women are shown blatantly offensive cases to annotate, it would be easier to align with their annotations due to the skewed sample of content they review. To mitigate these challenges, adopting a structured approach, such as Bayesian modeling, could clarify the influence of both observed and latent biases. Additionally, focusing analysis on instances where demographic groups diverge from the majority could reveal unique biases. Finally, we only use commercial LLMs like GPT4o and Gemini in our labeling since they are widely used, especially by interdisciplinary researchers~\cite{gilardi2023chatgpt}. In future, we will compare the results with comparable open source models like LLaMa.

Our study has several limitations that point to opportunities for future work. The analysis is restricted to the English language, which limits the applicability of our findings across languages with different cultural nuances and linguistic structures. Additionally, our focus on demographic factors like gender and race provides only a partial view of potential biases, possibly overlooking other demographic characteristics that may influence alignment patterns. % Our study is limited by the use of five datasets and two models, and while we acknowledge that additional demographics and data could be explored, we chose to stop after observing significant inconsistencies across these datasets. Given these inconsistencies, further expansion would not have addressed the core issue, though future work can continue to refine and build upon these findings. 
While we incorporated confounders such as document difficulty and annotator sensitivity, other factors like the target of offensive language, could further enhance our model. The rapid advancements in language models seek the need for refining evaluation methods to maintain relevance over time. Finally, we only assess the biases associated with default prompting, i.e., prompting without sociodemographic signals. While other researchers have looked into sociodemographic prompting~\cite{beck2023not,sun2023aligning,schäfer2024demographicsllmsdefaultannotation}, it is important to consider the default case in detail since in real-world settings, all relevant demographic variables may not be known a priori. Indeed, most recent work benchmarking the use of LLMs for content labeling do so without sociodemographic prompting~\cite{gilardi2023chatgpt,ziems2024can}.

\section{Ethics Statement}

% Essentializing (we assume that few annotators are stand-ins for their whole sociodemographic group)
% - limited conception of gender and race, and disregard for more complex backgrounds
% - focus on only english 
% - the potential negative effects of using LLMs as annotators (both for data workers and the quality of downstream models), considering that we are in part supporting this line of development

As language technologies become widely used for algorithmic decision-making, such as using NLP techniques for detecting offensive content as a type of content moderation tool, there are growing concerns about these technology's biases against marginalized populations; populations who are themselves most susceptible to receiving offensive attacks on platforms. In this work, we assess one aspect of such bias in the latest generation of language technology --- demographic misalignment in prompt-based Large Language Models. Our findings across multiple datasets show that current LLMs have varying and inconsistent alignment with different demographics, but have especially lower alignment with Black people, and for offensive and potentially offensive content.

While current work has assessed the utility of including demographic information in prompts to induce personas (`demographic steering')~\citep{santurkar2023whose}, also in the context of data annotation~\citep{sun2023aligning,beck2023not}, the preliminary results indicate that this type of steering does \textit{not} improve alignment. Therefore, we need to assess further strategies such as fine-tuning (instruction or otherwise), Retrieval Augmented Generation, or even pre-training to address this misalignment. 

We use three openly available datasets in our experiments, where the creators of these datasets made annotator demographics available along with the distribution of annotator labels~\citep{sun2023aligning,sap2020social,kennedy2020constructing}. All of the annotator data released by these authors are anonymized and we do not attempt to deanonymize any of the annotators. While we attempted to include understudied demographic identities in this work, particularly intersectional identities, we only consider men and women within our gender variables. This is because the annotations of other genders (non-binary people) were significantly fewer and could not be quantitatively modeled. However, it is important to represent gender minorities when assessing the alignment of LLMs, especially when they are used to label offensive content targeting these groups. We hope to address this in future work by having a larger and more diverse pool of annotators. 

While the results of our work indicate the need for strategies to improve alignment, there are also concerns of demographic essentialization and ecological fallacies~\citep{orlikowski2023ecological}; their demographic identity could be \textit{one} of the many factors affecting an annotator's perception of offensiveness. Other important factors to consider could be lived experiences, particularly past experiences with harassment. In future work, we hope to disentangle demographic and individual patterns when annotating content and devise ways of incorporating these into LLMs. 

\section{Reproducibility}
Our analysis relies on five distinct datasets, with four freely accessible through public repositories. The fifth dataset AwA (Annotators with Attitude) can be obtained through a formal request to the original authors. We used two models \textit{gpt-4o-mini-2024-07-18} and \textit{gemini-1.5-flash-002} both of which can be accessed from OpenAI and Google's APIs. The exact prompts that we used for all LLMs are included in the Appendix (Section~\ref{sec:prompts}). The code to run the experiments is available at \hyperlink{https://github.com/shayanalipour/llm-alignment-bias}{\textit{llm-demographic-bias}} repository on Github.

\bibliography{main}

\appendix

\section{Dataset Details and Demographic Distribution}\label{sec:app_data}

Here, we provided a more detailed summary of the 5 different datasets used in this work.

The \textbf{``Annotator with Attitudes''} (AwA) dataset~\cite{sap2021annotators} curates a dataset on potentially offensive content targeting Black people. We use their Breadth-of-Posts dataset, which contains 626 posts annotated by 177 annotators, totaling 3,349 annotations from different genders, ethnicities, and political backgrounds. Annotators were asked to rate how much they perceived each post as toxic, hateful, disrespectful, or offensive on a 5-point Likert scale, ranging from 1 (not at all) to 5 (very much so). To remain consistent across datasets, we use the annotators' gender and ethnicity as demographic variables.

UC Berkeley's \textbf{Measuring Hate Speech Corpus} (MHSC) ~\citep{kennedy2020constructing} contains 90,174 annotations from 7,725 annotators on 39,263 online comments. The metric used for comparison was the ``hatespeech'' ordinal label of each comment measuring the identified severity on a three-level scale: yes, no, and unclear.
%, gathered from annotators through a labeling survey task.

We used the \textbf{NLPositionality} (NLPos) dataset~\cite{santynlpositionality}, which was originally used for the hate speech detection task in their paper. This dataset contains annotations from 412 annotators on 299 posts, totaling 4,417 annotations. Annotators were asked to evaluate an instance using a 3-point scale.

% \subsection{POPQUORN}
The \textbf{POPQUORN dataset}~\citep{pei2023annotator} (POPQ) contains 12,088 annotations on 1,500 online comments from 243 annotators from a sample of the US adult population that was representative based on age, gender, and ethnicity. Annotators were asked to provide an offensiveness score of each text sample on a 1-5 scale, from ``Not offensive at all'' to ``Very offensive'', gathered from annotators through a multiple-choice task.

% \subsection{Social Bias Frames}
The \textbf{Social Bias Inference Corpus}~\citep{sap2020social}, referred to as SBIC, contains 109,349 annotations on 44,232 online posts from 280 annotators. The dataset is acknowledged by its creators to be racially skewed, with a vast majority of annotators being White and nearly none being both Black and male. Annotators were asked to evaluate whether each post could be considered offensive, disrespectful, or toxic to anyone/someone, with the following valid response options: 1 (Yes, this could be offensive), 2 (Maybe, I'm not sure), 3 (No, this is harmless), and 4 (I don't understand the post).

Naturally, a major factor determining the subjective offensiveness of a particular statement is the group or individual targeted by said statement. Therefore, we reasoned that demographic factors most frequently targeted by offensive statements were likely to have major effects on the perception of offensiveness, providing valuable insights in determining potential alignment. Considering available annotator data regarding targeted groups, it became clear that ethnicity and gender were the two most significant factors represented in targeted language. 11.7\% and 7.6\% of SBIC annotations noted targeted language towards Black folks and women respectively - figures around twice those corresponding to any other group. In MHSC, 35.7\% of annotations indicated offensive language targeting a racial group and 29.8\% targeted a gender, with 20.6\% targeting women and 16.9\% targeting Black people specifically. In contrast, age groups were targeted to a much lesser degree (1.5\% of Hate Speech annotations), displaying less evidence supporting its status as a largely influential factor. Overall, offensive language was shown to target racial and gender demographic groups, specifically Black people and women, indicating a high likelihood that annotator ethnicity and gender would have significant influences on the variable perception of offensive statements.

\section{Majority Results}
\label{sec:majority-results}
In this section, we present the results using the majority vote aggregation method to calculate the ground truth (e.g., offensiveness level). 

\begin{figure}[!htbp]
  \includegraphics[width=\linewidth]{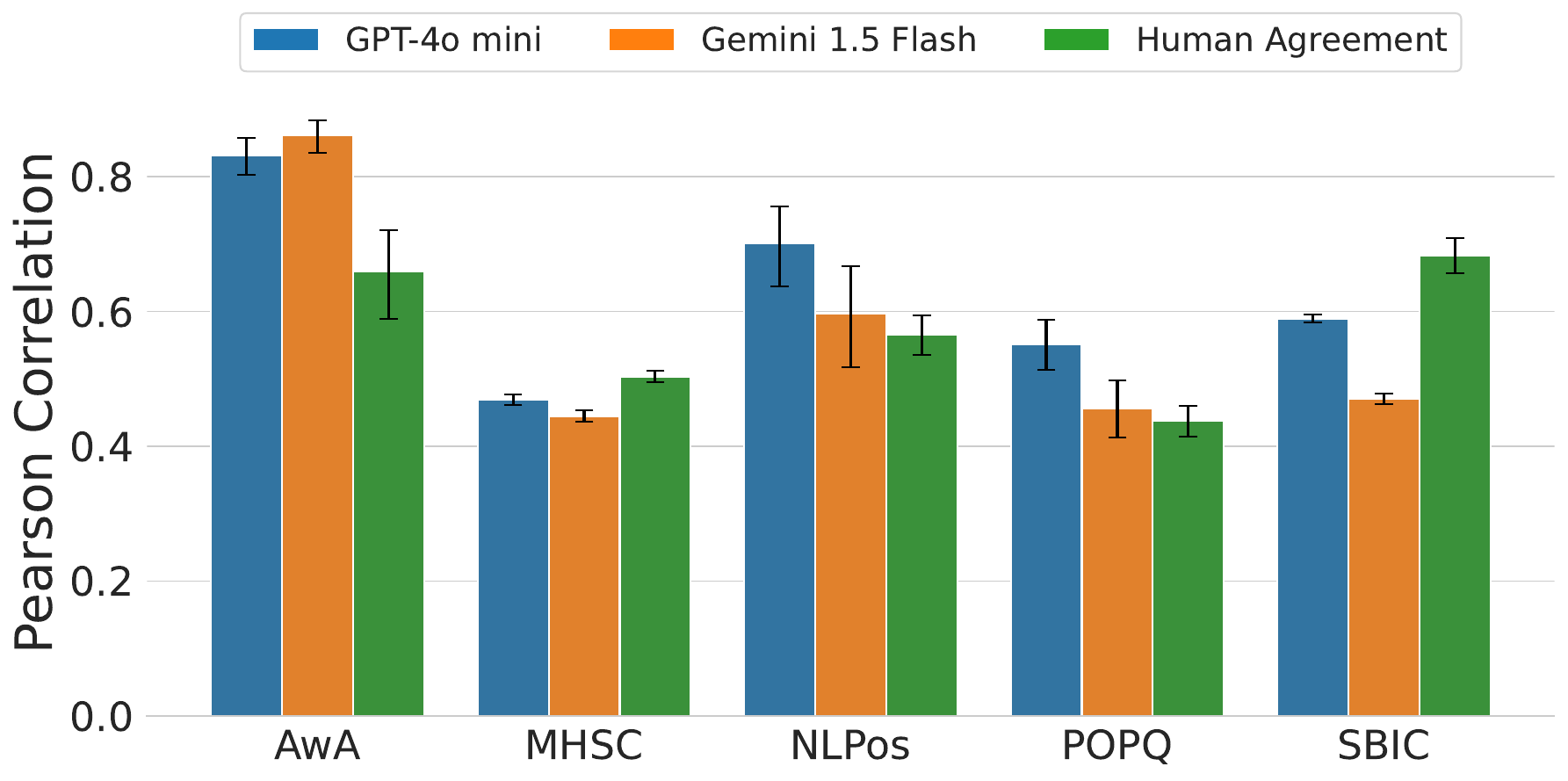}
\caption{
    Comparison of model correlations with human annotators against human agreement (individual annotators with their peers), highlighting how well models align with human judgment. The ground truth for each post is determined by the majority vote of annotators' labels. For human agreement, correlations are measured by leaving out one annotator and comparing their labels to the ground truth from the remaining annotators. Error bars represent 95\% confidence intervals.
}

    \label{fig:majority-human-vs-model}
\end{figure}

\textbf{RQ0}: Figure~\ref{fig:majority-human-vs-model} shows that the performance of language models on the subjective task of offensive language annotation aligns closely with the average performance of human annotators. This trend mirrors the pattern observed when the ground truth was calculated using rounded averages. Specifically, the correlations for GPT-4o mini range from  $r=0.47$ to $r=0.83$, while Gemini 1.5 Flash achieves correlations ranging from $r=0.44$ to $r=0.86$.

\begin{figure}[!htbp]
  \centering
  % \captionsetup[subfigure]{labelformat=empty}

  \begin{subfigure}{0.48\textwidth}
    \caption{Gender}
    \includegraphics[width=\linewidth]{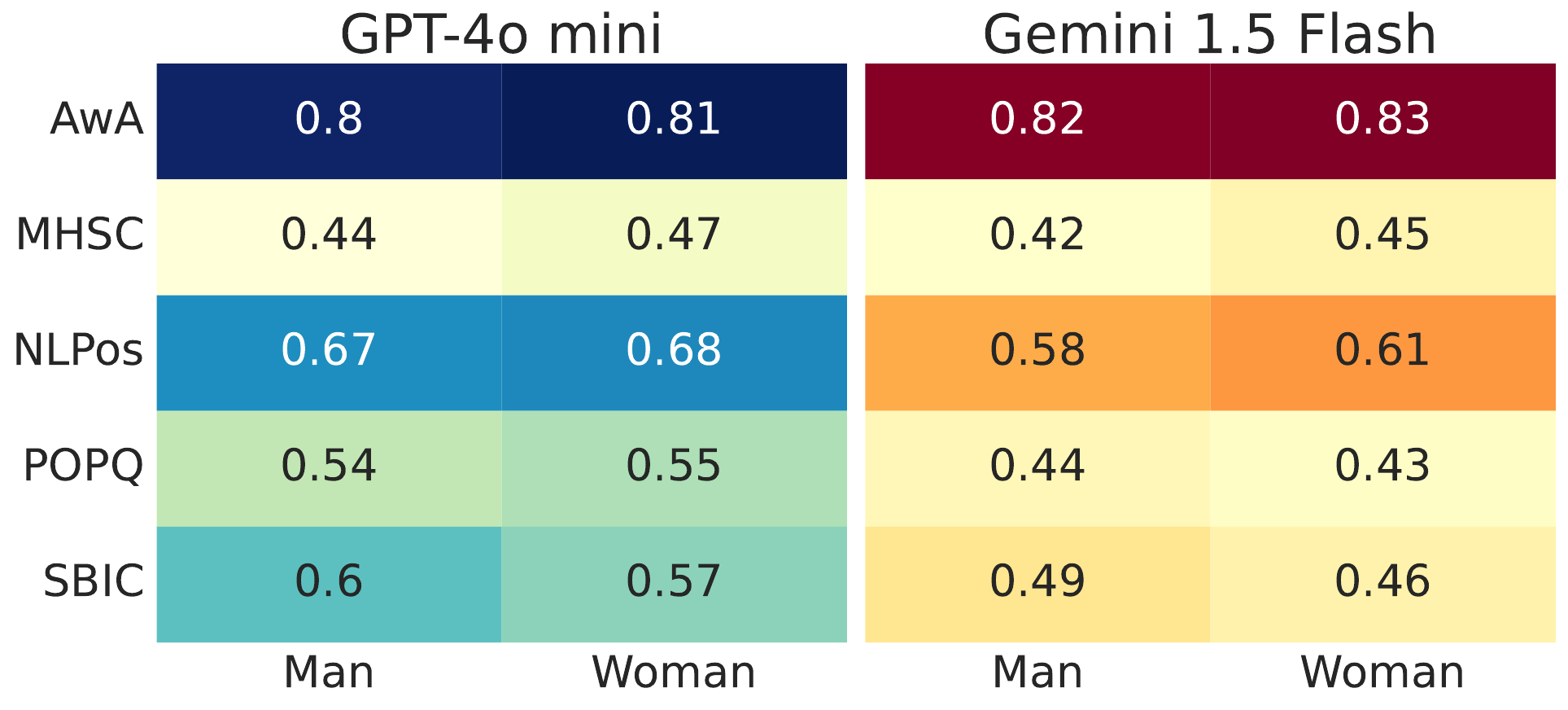}
  \end{subfigure}

  \vspace{0.5cm} 

  \begin{subfigure}{0.48\textwidth}
    \caption{Ethnicity}
    \includegraphics[width=\linewidth]{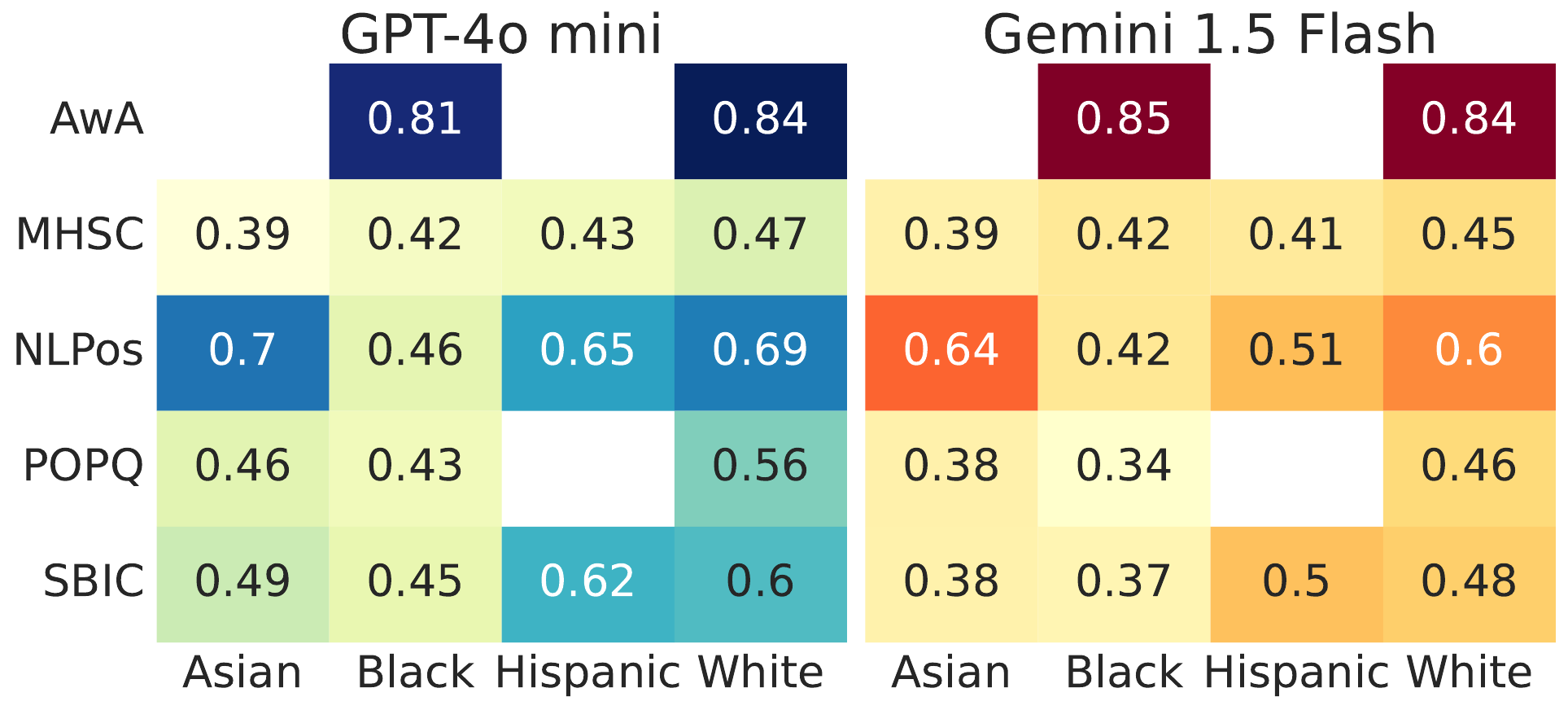}
  \end{subfigure}
  
    \caption{
        Pearson correlation coefficients between model outputs and human annotator labels, broken down by gender (a) and ethnicity (b) across five datasets. The ground truth for each post is determined by the majority vote of annotators from the target demographic. Darker shades indicate stronger correlations. Confidence intervals and p-values for statistical significance are reported in table \ref{tab:majority-demo-model-corr}.
    }
  \label{fig:gender-race-corrs-majority}
\end{figure}

\begin{figure}[!htbp]
  \centering

  \begin{subfigure}{0.48\textwidth}
    \caption{GPT-4o mini}
    \includegraphics[width=\linewidth]{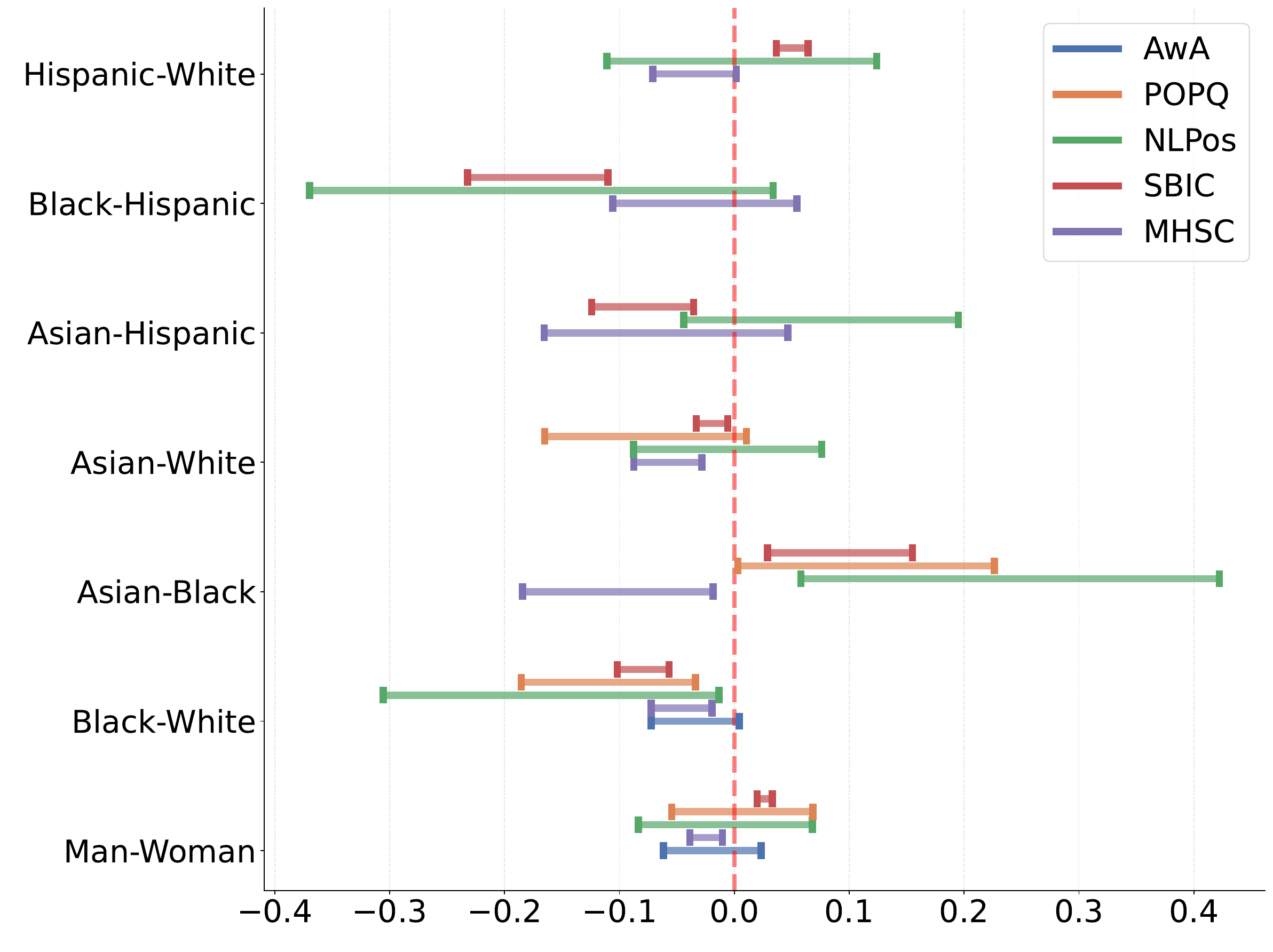}
  \end{subfigure}

  \vspace{0.5cm} 

  \begin{subfigure}{0.48\textwidth}
    \caption{Gemini 1.5 Flash}
    \includegraphics[width=\linewidth]{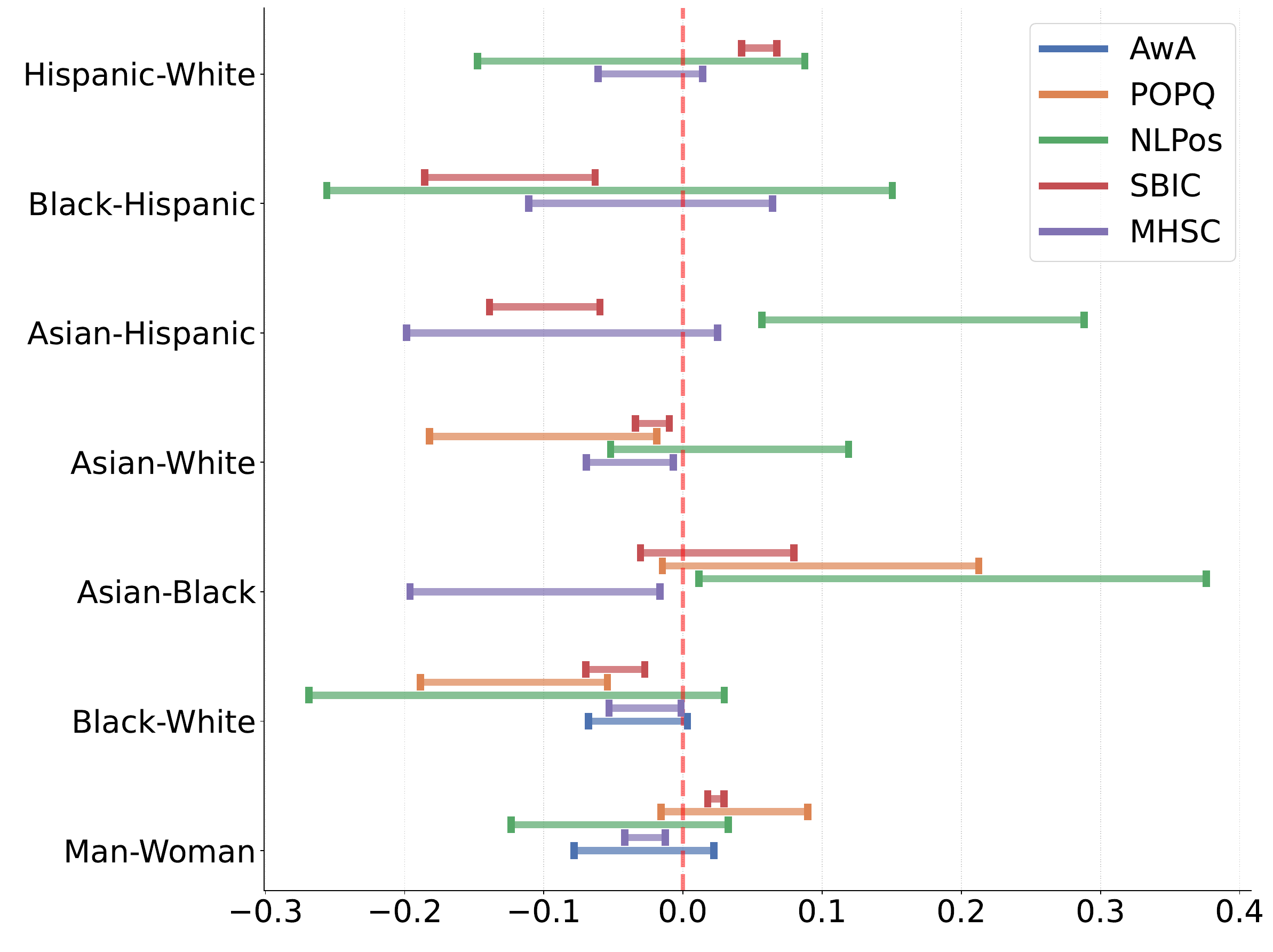}
  \end{subfigure}
    \caption{(Majority vote) The 95\% confidence intervals (CI) for the difference in correlation between the model's predictions and two demographic groups, computed as: $\Delta r = r(P, D_1) - r(P, D_2)$, where $P$ represents the model's predictions, and $D_1$ and $D_2$ are two demographic groups. Ground truth for each post is determined by the majority vote of annotators' labels. The intervals are derived from 1,000 bootstrap samples. If the CI includes zero, the difference is not statistically significant. See table~\ref{tab:majority-corr-diff} in the Appendix for further details.}
  \label{fig:majority_corr_diff}
\end{figure}

\textbf{ RQ1:} Figure~\ref{fig:gender-race-corrs-majority} presents the correlation values between LLM labels and the labels of annotators from each demographic group when using the majority vote aggregation method. The patterns observed largely align with those reported for the rounded average aggregation. For gender, in most datasets, both GPT-4o and Gemini align more closely with annotations from women. However, in addition to the SBIC database, we observe that in the POPQ dataset for the Gemini model, the correlation for men ($r=0.44 $) is also slightly higher than for women ($r=0.43$). For ethnicity, the majority vote results similarly show that model annotations often align better with the White demographic across datasets with some exceptions. For example, consistent with the rounded average results, in the SBIC dataset, models align better with the Hispanic demographic. In the NLPos dataset, we observe a reversal: Asian annotators achieve higher correlations compared to the White demographic ($ r=0.70$ vs. $ r=0.69 $ for GPT-4o mini, and $ r=0.64$ vs. $r=0.60$ for Gemini 1.5 Flash). Additionally, in the AwA dataset, Gemini shows a slightly higher correlation with the Black demographic ($ r=0.85 $) compared to the White demographic ($ r=0.84 $). As in the rounded average case, we assess the robustness of these differences using Steiger’s $ Z $ test and bootstrapping. Figure~\ref{fig:majority_corr_diff} indicate the same inconsistency patterns across demographic groups. For example, in the Gemini model, even the Black-White correlation differences show sign reversals at the tails of the 95\% confidence intervals for the AwA and NLPos datasets. Overall, Table~\ref{tab:majority-demo-model-corr} highlights that, while correlation values between models and demographic groups are statistically significant, the observed differences as evident in Table~\ref{tab:majority-corr-diff} are inconsistent. This inconsistency suggests that demographic identity alone does not consistently account for variations in model alignment, reaffirming the influence of other factors such as dataset composition or individual interpretation.

\begin{figure*}[!htbp]
  \centering

  \begin{subfigure}{\textwidth}
    \caption{GPT-4o mini}
    \includegraphics[width=\linewidth]{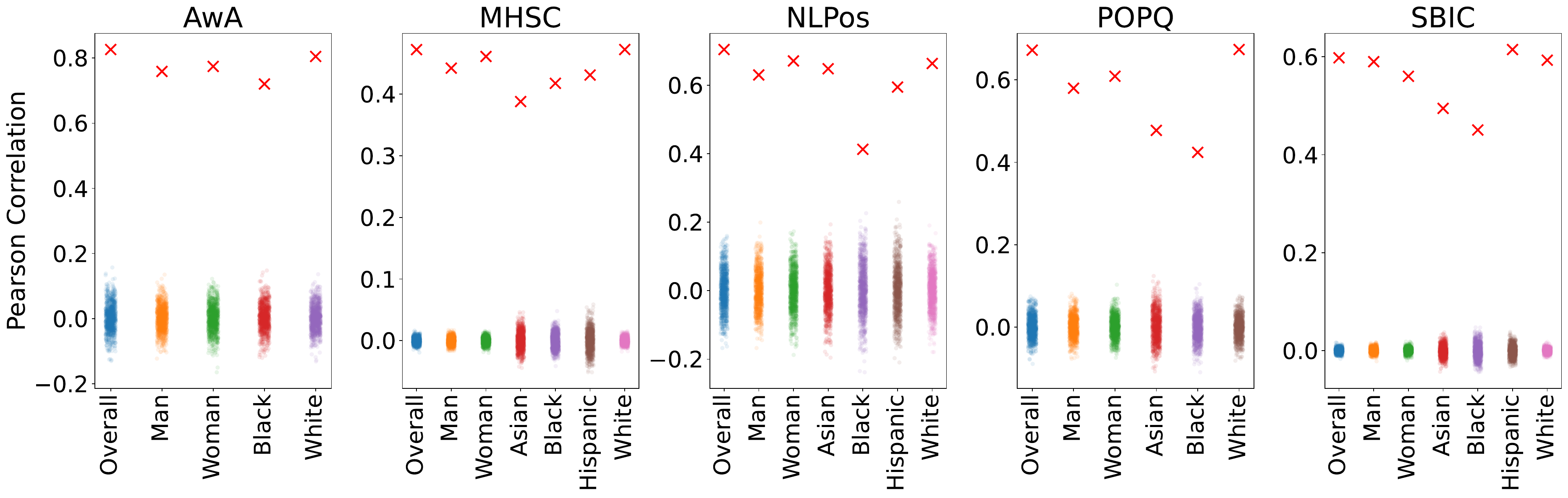}
  \end{subfigure}

  \vspace{0.5cm} 

  \begin{subfigure}{\textwidth}
    \caption{Gemini 1.5 Flash}
    \includegraphics[width=\linewidth]{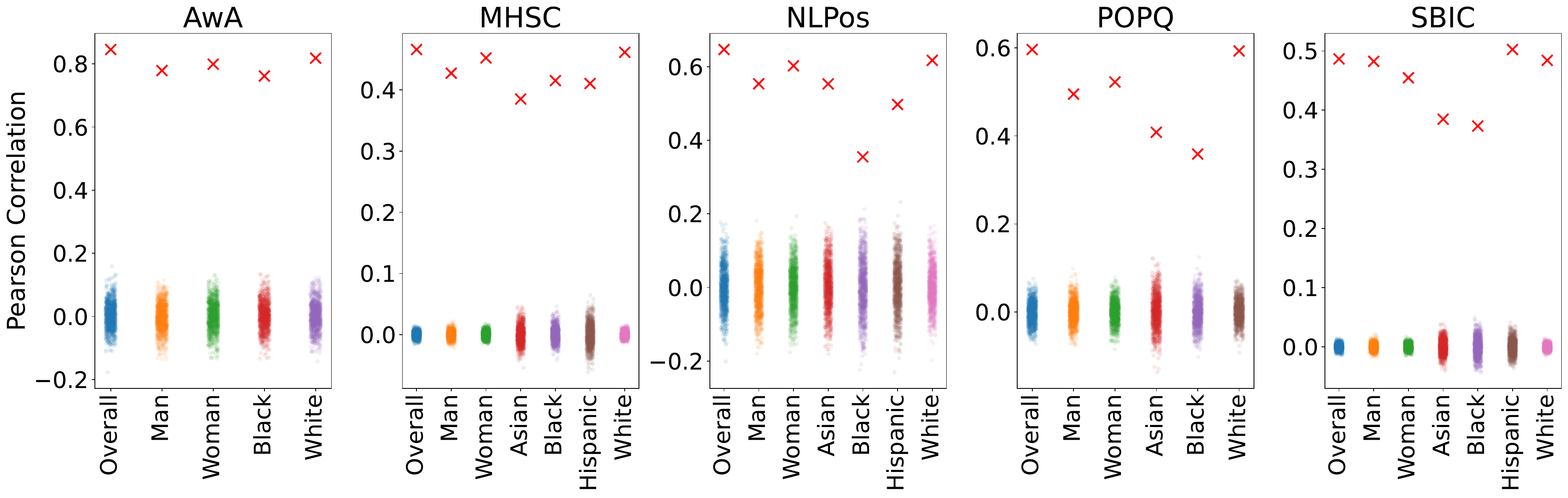}
  \end{subfigure}
  
\caption{(Average vote) Pearson correlation coefficients between model outputs and human annotator labels across demographic groups and datasets. Ground truth is determined by averaging the labels from annotators belonging to the target demographic. Red crosses show observed correlations, while colored scatter points represent the null distribution from 1,000 random label permutations. In both cases, observed correlations are consistently higher than shuffled ones, indicating statistical significance.}
    \label{fig:permutation-average}
\end{figure*}

\begin{figure*}[htbp]
  \centering

  \begin{subfigure}{\textwidth}
    \caption{GPT-4o mini}
    \includegraphics[width=\linewidth]{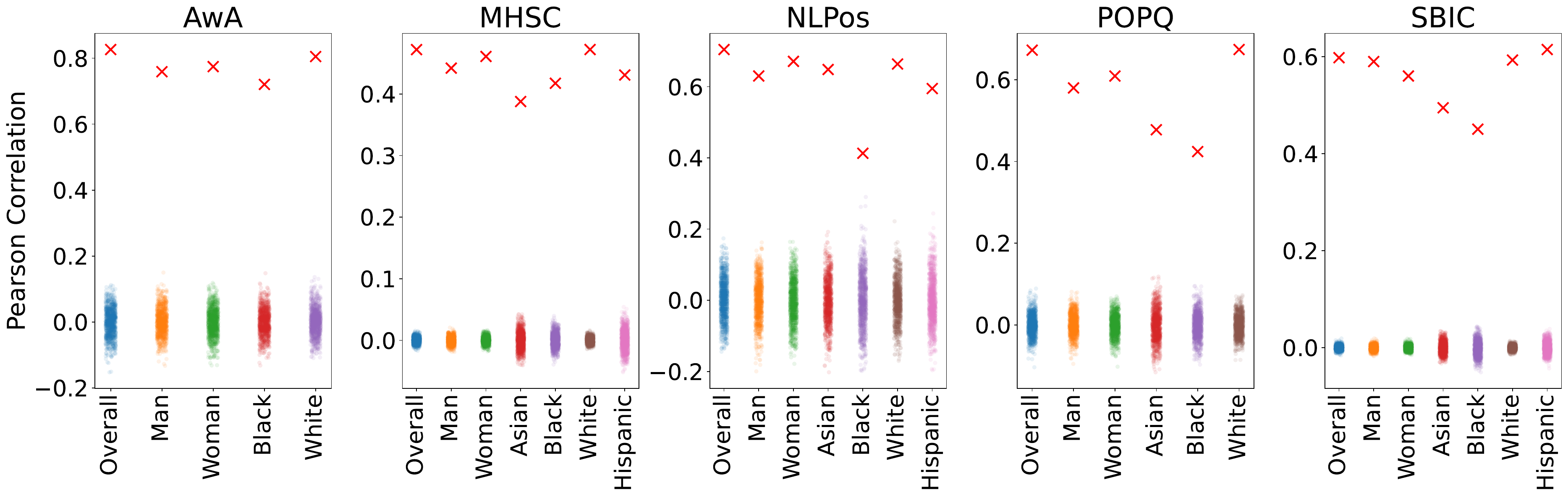}
  \end{subfigure}

  \vspace{0.5cm} 

  \begin{subfigure}{\textwidth}
    \caption{Gemini 1.5 Flash}
    \includegraphics[width=\linewidth]{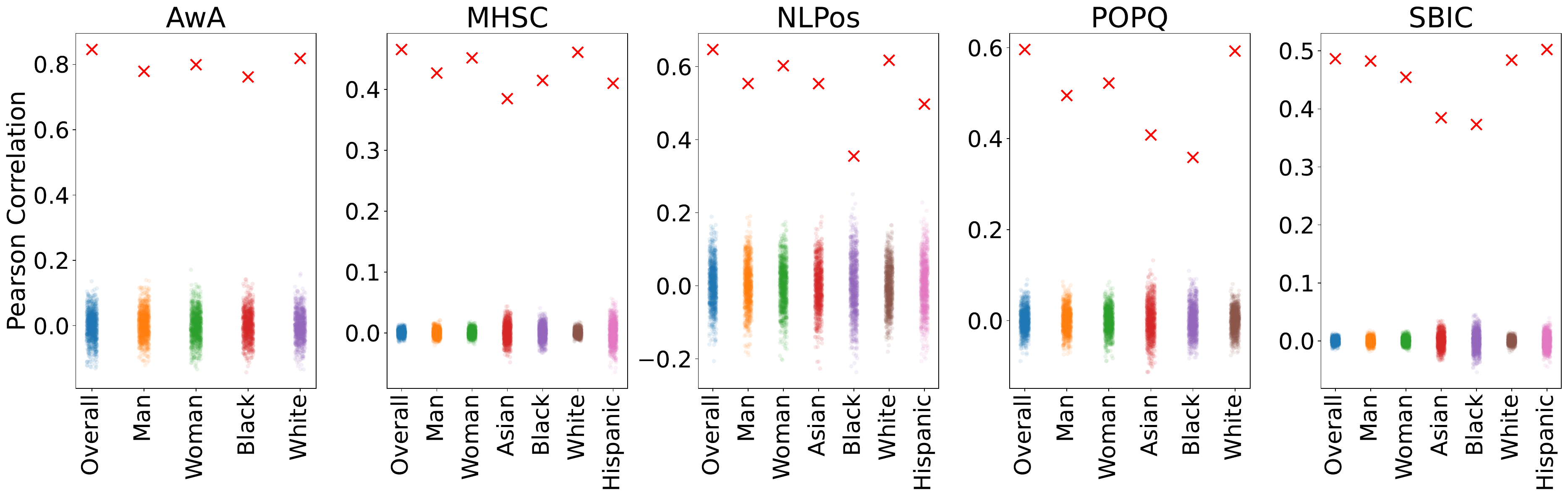}
  \end{subfigure}
\caption{(Majority vote) Pearson correlation coefficients between model outputs and human annotator labels across demographic groups and datasets. Ground truth is determined by the majority vote of annotators' labels. Red crosses show observed correlations, while colored scatter points represent the null distribution from 1,000 random label permutations. In both cases, observed correlations are consistently higher than shuffled ones, indicating statistical significance.}
\label{fig:permutation-majority}
\end{figure*}

\begin{table*}[!htbp]
\setlength{\tabcolsep}{3pt}
\footnotesize
\centering
\label{table:correlations-average}
\begin{tabular}{ll|cc|cc|cc|cc|cc}
\toprule
\multirow{3}{*}{\textbf{Demo.}} & \multirow{3}{*}{\textbf{Model}} & \multicolumn{2}{c}{\textbf{AwA}} & \multicolumn{2}{c}{\textbf{MHSC}} & \multicolumn{2}{c}{\textbf{NLPos}} & \multicolumn{2}{c}{\textbf{POPQ}} & \multicolumn{2}{c}{\textbf{SBIC}}\\
\cmidrule(lr){3-12}
& & Corr. & 95\% CI & Corr. & 95\% CI & Corr. & 95\% CI & Corr. & 95\% CI & Corr. & 95\% CI \\
\midrule
\multirow{2}{*}{Overall} & GPT & $0.83^{*}$ & $(0.80, 0.85)$ & $0.47^{*}$ & $(0.46, 0.48)$ & $0.70^{*}$ & $(0.64, 0.76)$ & $0.67^{*}$ & $(0.64, 0.70)$ & $0.60^{*}$ & $(0.59, 0.60)$ \\
& Gemini & $0.85^{*}$ & $(0.82, 0.87)$ & $0.47^{*}$ & $(0.46, 0.47)$ & $0.65^{*}$ & $(0.57, 0.71)$ & $0.60^{*}$ & $(0.56, 0.63)$ & $0.49^{*}$ & $(0.48, 0.49)$ \\
\midrule
\multirow{2}{*}{Man} & GPT & $0.76^{*}$ & $(0.72, 0.79)$ & $0.44^{*}$ & $(0.43, 0.45)$ & $0.63^{*}$ & $(0.56, 0.69)$ & $0.58^{*}$ & $(0.55, 0.61)$ & $0.59^{*}$ & $(0.58, 0.60)$ \\
& Gemini & $0.78^{*}$ & $(0.74, 0.81)$ & $0.43^{*}$ & $(0.42, 0.44)$ & $0.55^{*}$ & $(0.47, 0.63)$ & $0.49^{*}$ & $(0.46, 0.53)$ & $0.48^{*}$ & $(0.47, 0.49)$ \\
\midrule
\multirow{2}{*}{Woman} & GPT & $0.77^{*}$ & $(0.74, 0.80)$ & $0.46^{*}$ & $(0.45, 0.47)$ & $0.67^{*}$ & $(0.60, 0.73)$ & $0.61^{*}$ & $(0.58, 0.64)$ & $0.56^{*}$ & $(0.55, 0.57)$ \\
& Gemini & $0.80^{*}$ & $(0.77, 0.83)$ & $0.45^{*}$ & $(0.44, 0.46)$ & $0.60^{*}$ & $(0.52, 0.67)$ & $0.52^{*}$ & $(0.48, 0.56)$ & $0.45^{*}$ & $(0.45, 0.46)$ \\
\midrule
\multirow{2}{*}{Asian} & GPT & -- & -- & $0.39^{*}$ & $(0.36, 0.41)$ & $0.65^{*}$ & $(0.57, 0.71)$ & $0.48^{*}$ & $(0.42, 0.53)$ & $0.49^{*}$ & $(0.48, 0.51)$ \\
& Gemini & -- & -- & $0.38^{*}$ & $(0.36, 0.41)$ & $0.55^{*}$ & $(0.46, 0.63)$ & $0.41^{*}$ & $(0.35, 0.47)$ & $0.38^{*}$ & $(0.36, 0.41)$ \\
\midrule
\multirow{2}{*}{Black} & GPT & $0.72^{*}$ & $(0.68, 0.76)$ & $0.42^{*}$ & $(0.40, 0.44)$ & $0.41^{*}$ & $(0.28, 0.53)$ & $0.42^{*}$ & $(0.37, 0.47)$ & $0.45^{*}$ & $(0.43, 0.47)$ \\
& Gemini & $0.76^{*}$ & $(0.72, 0.79)$ & $0.41^{*}$ & $(0.40, 0.43)$ & $0.35^{*}$ & $(0.22, 0.48)$ & $0.36^{*}$ & $(0.30, 0.41)$ & $0.37^{*}$ & $(0.35, 0.40)$ \\
\midrule
\multirow{2}{*}{Hispanic} & GPT & -- & -- & $0.43^{*}$ & $(0.40, 0.46)$ & $0.59^{*}$ & $(0.50, 0.67)$ & -- & -- & $0.61^{*}$ & $(0.60, 0.63)$ \\
& Gemini & -- & -- & $0.41^{*}$ & $(0.38, 0.44)$ & $0.50^{*}$ & $(0.39, 0.59)$ & -- & -- & $0.50^{*}$ & $(0.48, 0.52)$ \\
\midrule
\multirow{2}{*}{White} & GPT & $0.80^{*}$ & $(0.78, 0.83)$ & $0.47^{*}$ & $(0.46, 0.48)$ & $0.66^{*}$ & $(0.59, 0.72)$ & $0.67^{*}$ & $(0.65, 0.70)$ & $0.59^{*}$ & $(0.59, 0.60)$ \\
& Gemini & $0.82^{*}$ & $(0.79, 0.84)$ & $0.46^{*}$ & $(0.45, 0.47)$ & $0.62^{*}$ & $(0.54, 0.68)$ & $0.59^{*}$ & $(0.56, 0.62)$ & $0.48^{*}$ & $(0.48, 0.49)$ \\

\bottomrule
\end{tabular}
\caption{(Average vote) Correlation results across datasets, models, and demographics, with ground truth determined by averaging labels from annotators within the target demographic. $^{*}$ indicates statistical significance at $p$-value < 0.05 (corrected for multiple comparisons).}
\label{tab:average-demo-model-corr}
\end{table*}

\begin{table*}[!htbp]
\setlength{\tabcolsep}{3pt}
\footnotesize
\centering
\label{table:correlations-majority}
\begin{tabular}{ll|cc|cc|cc|cc|cc}
\toprule
\multirow{3}{*}{\textbf{Demo.}} & \multirow{3}{*}{\textbf{Model}} & \multicolumn{2}{c}{\textbf{AwA}} & \multicolumn{2}{c}{\textbf{MHSC}} & \multicolumn{2}{c}{\textbf{NLPos}} & \multicolumn{2}{c}{\textbf{POPQ}} & \multicolumn{2}{c}{\textbf{SBIC}}\\
\cmidrule(lr){3-12}
& & Corr. & 95\% CI & Corr. & 95\% CI & Corr. & 95\% CI & Corr. & 95\% CI & Corr. & 95\% CI \\
\midrule
\multirow{2}{*}{Overall} & GPT & $0.83^{*}$ & $(0.80, 0.86)$ & $0.47^{*}$ & $(0.46, 0.48)$ & $0.70^{*}$ & $(0.64, 0.76)$ & $0.55^{*}$ & $(0.51, 0.59)$ & $0.59^{*}$ & $(0.58, 0.60)$ \\
& Gemini & $0.86^{*}$ & $(0.84, 0.88)$ & $0.44^{*}$ & $(0.44, 0.45)$ & $0.60^{*}$ & $(0.52, 0.67)$ & $0.46^{*}$ & $(0.41, 0.50)$ & $0.47^{*}$ & $(0.46, 0.48)$ \\
\midrule
\multirow{2}{*}{Man} & GPT & $0.80^{*}$ & $(0.76, 0.83)$ & $0.44^{*}$ & $(0.43, 0.45)$ & $0.67^{*}$ & $(0.60, 0.73)$ & $0.54^{*}$ & $(0.50, 0.58)$ & $0.60^{*}$ & $(0.60, 0.61)$ \\
& Gemini & $0.82^{*}$ & $(0.79, 0.85)$ & $0.42^{*}$ & $(0.41, 0.43)$ & $0.58^{*}$ & $(0.50, 0.65)$ & $0.44^{*}$ & $(0.40, 0.49)$ & $0.49^{*}$ & $(0.48, 0.50)$ \\
\midrule
\multirow{2}{*}{Woman} & GPT & $0.81^{*}$ & $(0.77, 0.84)$ & $0.47^{*}$ & $(0.46, 0.48)$ & $0.68^{*}$ & $(0.61, 0.74)$ & $0.55^{*}$ & $(0.51, 0.59)$ & $0.57^{*}$ & $(0.56, 0.58)$ \\
& Gemini & $0.83^{*}$ & $(0.80, 0.86)$ & $0.45^{*}$ & $(0.44, 0.46)$ & $0.61^{*}$ & $(0.53, 0.68)$ & $0.43^{*}$ & $(0.38, 0.47)$ & $0.46^{*}$ & $(0.45, 0.47)$ \\
\midrule
\multirow{2}{*}{Asian} & GPT & -- & -- & $0.39^{*}$ & $(0.37, 0.41)$ & $0.70^{*}$ & $(0.63, 0.76)$ & $0.46^{*}$ & $(0.40, 0.52)$ & $0.49^{*}$ & $(0.48, 0.51)$ \\
& Gemini & -- & -- & $0.39^{*}$ & $(0.36, 0.41)$ & $0.64^{*}$ & $(0.55, 0.71)$ & $0.38^{*}$ & $(0.32, 0.45)$ & $0.38^{*}$ & $(0.36, 0.40)$ \\
\midrule
\multirow{2}{*}{Black} & GPT & $0.81^{*}$ & $(0.77, 0.84)$ & $0.42^{*}$ & $(0.40, 0.44)$ & $0.46^{*}$ & $(0.32, 0.58)$ & $0.43^{*}$ & $(0.37, 0.49)$ & $0.45^{*}$ & $(0.43, 0.48)$ \\
& Gemini & $0.85^{*}$ & $(0.81, 0.88)$ & $0.42^{*}$ & $(0.40, 0.44)$ & $0.42^{*}$ & $(0.28, 0.54)$ & $0.34^{*}$ & $(0.27, 0.40)$ & $0.37^{*}$ & $(0.35, 0.40)$ \\
\midrule
\multirow{2}{*}{Hispanic} & GPT & -- & -- & $0.43^{*}$ & $(0.40, 0.46)$ & $0.65^{*}$ & $(0.56, 0.72)$ & -- & -- & $0.62^{*}$ & $(0.60, 0.63)$ \\
& Gemini & -- & -- & $0.41^{*}$ & $(0.38, 0.44)$ & $0.51^{*}$ & $(0.40, 0.61)$ & -- & -- & $0.50^{*}$ & $(0.49, 0.52)$ \\
\midrule
\multirow{2}{*}{White} & GPT & $0.84^{*}$ & $(0.81, 0.86)$ & $0.47^{*}$ & $(0.46, 0.48)$ & $0.69^{*}$ & $(0.62, 0.75)$ & $0.56^{*}$ & $(0.52, 0.59)$ & $0.60^{*}$ & $(0.59, 0.61)$ \\
& Gemini & $0.84^{*}$ & $(0.81, 0.87)$ & $0.45^{*}$ & $(0.44, 0.46)$ & $0.60^{*}$ & $(0.51, 0.67)$ & $0.46^{*}$ & $(0.42, 0.50)$ & $0.48^{*}$ & $(0.47, 0.49)$ \\

\bottomrule
\end{tabular}
\caption{(Majority vote) Correlation results across datasets, models, and demographics, with ground truth determined by the majority vote of annotators’ labels. $^{*}$ indicates statistical significance at $p$-value < 0.05 (corrected for multiple comparisons).}
\label{tab:majority-demo-model-corr}
\end{table*}

\begin{table*}[!htbp]
\centering
\footnotesize
\setlength{\tabcolsep}{5pt}
\begin{tabular}{ll|c|c|c|c|c}
\toprule
{\textbf{Demo. Pair}} & {\textbf{Model}} & \multicolumn{1}{|c}{\textbf{AwA}} & \multicolumn{1}{|c}{\textbf{MHSC}} & \multicolumn{1}{|c}{\textbf{NLPos}} & \multicolumn{1}{|c}{\textbf{POPQ}} & \multicolumn{1}{|c}{\textbf{SBIC}}\\
\midrule
\multirow{2}{*}{Asian - Black} & GPT & -- & (-0.18, -0.01)$^{**}$ & (0.09, 0.38)$^{**}$ & (0.03, 0.23)$^{**}$ & (0.04, 0.16)$^{**}$ \\
& Gemini & -- & (-0.18, -0.01)$^{*}$ & (0.08, 0.39)$^{**}$ & (0.00, 0.19)$^{*}$ & (-0.02, 0.09) \\
\midrule
\multirow{2}{*}{Asian - Hispanic} & GPT & -- & (-0.16, 0.03) & (-0.01, 0.20) & -- & (-0.12, -0.03)$^{**}$ \\
& Gemini & -- & (-0.19, 0.03) & (0.03, 0.26)$^{**}$ & -- & (-0.14, -0.06)$^{**}$ \\
\midrule
\multirow{2}{*}{Asian - White} & GPT & -- & (-0.11, -0.05)$^{***}$ & (-0.10, 0.06) & (-0.27, -0.14)$^{***}$ & (-0.07, -0.05)$^{***}$ \\
& Gemini & -- & (-0.09, -0.02)$^{***}$ & (-0.15, 0.02) & (-0.26, -0.13)$^{***}$ & (-0.08, -0.05)$^{***}$ \\
\midrule
\multirow{2}{*}{Black - Hispanic} & GPT & -- & (-0.11, 0.05) & (-0.36, -0.02)$^{*}$ & -- & (-0.25, -0.12)$^{***}$ \\
& Gemini & -- & (-0.12, 0.06) & (-0.34, 0.01) & -- & (-0.19, -0.08)$^{***}$ \\
\midrule
\multirow{2}{*}{Black - White} & GPT & (-0.12, -0.05)$^{***}$ & (-0.09, -0.04)$^{***}$ & (-0.34, -0.09)$^{**}$ & (-0.29, -0.19)$^{***}$ & (-0.13, -0.09)$^{***}$ \\
& Gemini & (-0.10, -0.01)$^{**}$ & (-0.08, -0.03)$^{***}$ & (-0.38, -0.10)$^{***}$ & (-0.29, -0.18)$^{***}$ & (-0.10, -0.06)$^{***}$ \\
\midrule
\multirow{2}{*}{Hispanic - White} & GPT & -- & (-0.08, -0.01)$^{**}$ & (-0.15, 0.07) & -- & (0.02, 0.04)$^{***}$ \\
& Gemini & -- & (-0.08, -0.01)$^{*}$ & (-0.23, -0.00)$^{*}$ & -- & (0.02, 0.04)$^{**}$ \\
\midrule
\multirow{2}{*}{Man - Woman} & GPT & (-0.06, 0.02) & (-0.04, -0.01)$^{***}$ & (-0.11, 0.03) & (-0.07, 0.01) & (0.02, 0.03)$^{***}$ \\
& Gemini & (-0.07, 0.01) & (-0.04, -0.01)$^{***}$ & (-0.13, 0.03) & (-0.07, 0.02) & (0.02, 0.03)$^{***}$ \\

\bottomrule
\end{tabular}
\caption{(Average vote) The 95\% confidence intervals (CI) for the difference in correlation between the model's predictions and two demographic groups, computed as: $\Delta r = r(P, D_1) - r(P, D_2)$, where $P$ represents the model's predictions, and $D_1$ and $D_2$ are two demographic groups. Ground truth for each post is determined by averaging the labels from annotators in the target demographic. The CIs are based on 1000 bootstrap samples. $p$-values were computed using Steiger's Z test and corrected for multiple comparisons with Holm's method. Significance levels are indicated by asterisks: *$p$ < 0.1, **$p$ < 0.05, ***$p$ < 0.001.}
\label{tab:average-corr-diff}
\end{table*}

\begin{table*}[!htbp]
\centering
\footnotesize
\setlength{\tabcolsep}{5pt}
\begin{tabular}{ll|c|c|c|c|c}
\toprule
{\textbf{Demo. Pair}} & {\textbf{Model}} & \multicolumn{1}{|c}{\textbf{AwA}} & \multicolumn{1}{|c}{\textbf{MHSC}} & \multicolumn{1}{|c}{\textbf{NLPos}} & \multicolumn{1}{|c}{\textbf{POPQ}} & \multicolumn{1}{|c}{\textbf{SBIC}}\\
\midrule
\multirow{2}{*}{Asian - Black} & GPT & -- & (-0.18, -0.02)$^{**}$ & (0.06, 0.42)$^{**}$ & (0.00, 0.23) & (0.03, 0.16)$^{**}$ \\
& Gemini & -- & (-0.20, -0.02)$^{*}$ & (0.01, 0.38)$^{*}$ & (-0.01, 0.21) & (-0.03, 0.08) \\
\midrule
\multirow{2}{*}{Asian - Hispanic} & GPT & -- & (-0.17, 0.05) & (-0.04, 0.20) & -- & (-0.12, -0.04)$^{**}$ \\
& Gemini & -- & (-0.20, 0.02) & (0.06, 0.29)$^{**}$ & -- & (-0.14, -0.06)$^{**}$ \\
\midrule
\multirow{2}{*}{Asian - White} & GPT & -- & (-0.09, -0.03)$^{***}$ & (-0.09, 0.08) & (-0.16, 0.01) & (-0.03, -0.01)$^{**}$ \\
& Gemini & -- & (-0.07, -0.01)$^{*}$ & (-0.05, 0.12) & (-0.18, -0.02)$^{*}$ & (-0.03, -0.01)$^{**}$ \\
\midrule
\multirow{2}{*}{Black - Hispanic} & GPT & -- & (-0.11, 0.05) & (-0.37, 0.03) & -- & (-0.23, -0.11)$^{***}$ \\
& Gemini & -- & (-0.11, 0.06) & (-0.26, 0.15) & -- & (-0.19, -0.06)$^{**}$ \\
\midrule
\multirow{2}{*}{Black - White} & GPT & (-0.07, 0.00)$^{**}$ & (-0.07, -0.02)$^{***}$ & (-0.31, -0.01)$^{**}$ & (-0.19, -0.03)$^{**}$ & (-0.10, -0.06)$^{***}$ \\
& Gemini & (-0.07, 0.00)$^{**}$ & (-0.05, -0.00)$^{*}$ & (-0.27, 0.03) & (-0.19, -0.05)$^{**}$ & (-0.07, -0.03)$^{**}$ \\
\midrule
\multirow{2}{*}{Hispanic - White} & GPT & -- & (-0.07, 0.00) & (-0.11, 0.12) & -- & (0.04, 0.06)$^{***}$ \\
& Gemini & -- & (-0.06, 0.01) & (-0.15, 0.09) & -- & (0.04, 0.07)$^{***}$ \\
\midrule
\multirow{2}{*}{Man - Woman} & GPT & (-0.06, 0.02) & (-0.04, -0.01)$^{**}$ & (-0.08, 0.07) & (-0.05, 0.07) & (0.02, 0.03)$^{***}$ \\
& Gemini & (-0.08, 0.02)$^{*}$ & (-0.04, -0.01)$^{***}$ & (-0.12, 0.03) & (-0.02, 0.09) & (0.02, 0.03)$^{***}$ \\

\bottomrule
\end{tabular}
\caption{(Majority vote) The 95\% confidence intervals (CI) for the difference in correlation between the model's predictions and two demographic groups, computed as: $\Delta r = r(P, D_1) - r(P, D_2)$, where $P$ represents the model's predictions, and $D_1$ and $D_2$ are two demographic groups. Ground truth for each post is determined by the majority vote of annotators' labels. The CIs are based on 1000 bootstrap samples. $p$-values were computed using Steiger's Z test and corrected for multiple comparisons with Holm's method. Significance levels are indicated by asterisks: *$p$ < 0.1, **$p$ < 0.05, ***$p$ < 0.001.}
\label{tab:majority-corr-diff}
\end{table*}

\section{Prompts}
\label{sec:prompts}

\begin{table*}[hb]
\footnotesize
\centering
\setlength{\tabcolsep}{4pt}
\begin{tabular}[t]{lp{0.9\textwidth}} % Set the width for the second column
\toprule
Dataset & Prompt \\
\midrule
AwA & How much could this post be seen as toxic/hateful/disrespectful or offensive to anyone, on a scale from 1 (not at all) to 5 (very much so)? Valid options: 1, 2, 3, 4, 5. Begin your response by selecting one of the valid options, then provide a concise explanation for your rating. \\
\midrule
MHSC & Does this comment contain hate speech, defined as “bias-motivated, hostile and malicious language targeted at a person/group because of their actual or perceived innate characteristics, especially when the group is unnecessarily labeled (e.g., “jew bankers”, “nâĂŞ hooligans”).” The comment contains hate speech if slurs or derogatory language are used toward any person or group regardless of whether or not the target is a member of the slur-target group. (e.g. a comment calling Trump a faggot or a n***** still contains hate speech). Valid options are 1 (Yes), 2 (Unclear), 3 (No). Begin your response by mentioning one of the valid options, then provide a concise explanation for your rating.\\ \midrule
NLPos & Do you think this post is hateful speech? Valid options are 1 (Yes), 2 (Not sure), 3 (No). Begin your response by mentioning one of the valid options, then provide a concise explanation for your rating. \\
\midrule
POPQ & Consider you read the above comment on Reddit. How offensive do you think it is, on a scale from 1 (not offensive at all) to 5 (very offensive)? Valid options: 1, 2, 3, 4, 5. Begin your response by selecting one of the valid options, then provide a concise explanation for your rating.\\ \midrule
SBIC & Could this post be considered offensive, disrespectful, or toxic to anyone/someone? Valid options are: 1 (Yes, this could be offensive), 2 (Maybe, I'm not sure), 3 (No, this is harmless), 4 (I don't understand the post). Begin your response by mentioning one of the valid options, then provide a concise explanation for your rating. \\
\bottomrule
\end{tabular}
\caption{Prompts used for inference to annotate comments and posts, based on the original questions and wording provided to human annotators in each dataset.}
\label{tab:prompts}
\end{table*}

\end{document}